\documentclass[journal]{IEEEtran}

\usepackage{amsfonts}
\usepackage{rotating}
\usepackage{graphicx}
\usepackage{epstopdf}
\usepackage{color}
\usepackage{bm}
\usepackage{amsmath}
\usepackage{subfigure}
\usepackage[linewidth=1pt]{mdframed}
\usepackage{lipsum}
\usepackage{cite}
\usepackage{extarrows}
\usepackage{subfigure}
\ifCLASSINFOpdf
\else
\fi

\def\^{\hat}
\def\~{\tilde}

\def\3h{{3\over 2}}

\def\eqn#1$${\eqno{{\rm #1}}$$}

\def\~{\tilde}

\def\^{\hat}

\def\XXint#1#2#3{{\setbox0=\hbox{$#1{#2#3}{\int}$}
     \vcenter{\hbox{$#2#3$}}\kern-.5\wd0}}

\begin{document}
%
\title{Spectral Numerical Mode Matching Method for 3D Layered Multi-Region Structures}
%
%
%

\author{Jie~Liu\IEEEmembership{},
        Na~Liu\IEEEmembership{},
        and~Qing~Huo~Liu,~\IEEEmembership{Fellow,~IEEE} 

\thanks{This research is partially supported by the National Key R\&D Program of China under Grant 2018YFC0603503, and in part by the National Natural Science Foundation of China under Grants 61871340, 11604276 and 61871462, and in part by the Ph.D. Start-up Fund of the Natural Science Foundation of Guangdong Province, China, under Grant 2016A030310372.
 \emph{(Corresponding author: Qing Huo Liu)}}

\thanks{J. Liu is with Institute of Electromagnetics and Acoustics, and Department of Electronic Science, Xiamen University, Xiamen 361005, China (e-mail: liujie190484@163.com}

\thanks{N. Liu is with the Institute of Electromagnetics and Acoustics, Xiamen University, Xiamen 361005, China, and also with the Shenzhen Research Institute, Xiamen University, Shenzhen 518057, China (e-mail: liuna@xmu.edu.cn)}

\thanks{Q. H. Liu is with the Department of Electrical and Computer Engineering,
Duke University, Durham, NC 27708 USA (e-mail: qhliu@duke.edu).}


}

\maketitle
\begin{abstract} The spectral numerical mode-matching (SNMM) method is developed to simulate the 3D layered multi-region structures. The SNMM method is a semi-analytical solver having the properties of dimensionality reduction to reduce computational costs; it is especially useful for microwave and optical integrated circuits where fabrication is often done in a layered structure. Furthermore, at some layer interfaces, very thin surfaces such as good conductor surfaces and metasurfaces can be deposited to achieve desired properties such as high absorbance and/or anomalous reflection/refraction. In this work, the 3D SNMM method is further extended from a single interface to multiple layers so that the electromagnetic propagation and scattering in the longitudinal direction is treated analytically through reflection and transmission matrices by using the eigenmode expansions in the transverse directions. Therefore, it effectively reduces the original 3D problem into a series of 2D eigenvalue problems for periodic structures. We apply this method to characterize metasurfaces and lithography models, and show that the SNMM method is especially efficient when the longitudinal layer thicknesses are large compared with wavelength. Numerical experiments indicate that the SNMM method is highly efficient and accurate for the metasurfaces and the lithography models.
\end{abstract}
\begin{IEEEkeywords}
Bloch (Floquet) periodic eigenmodes, lithography, metasurface, mixed spectral element method, spectral numerical mode-matching method.
\end{IEEEkeywords}

%

\IEEEpeerreviewmaketitle

\section{Introduction}

\begin{figure}[!t]
  \centering
  {
   \includegraphics[width=1.0\columnwidth,draft=false]{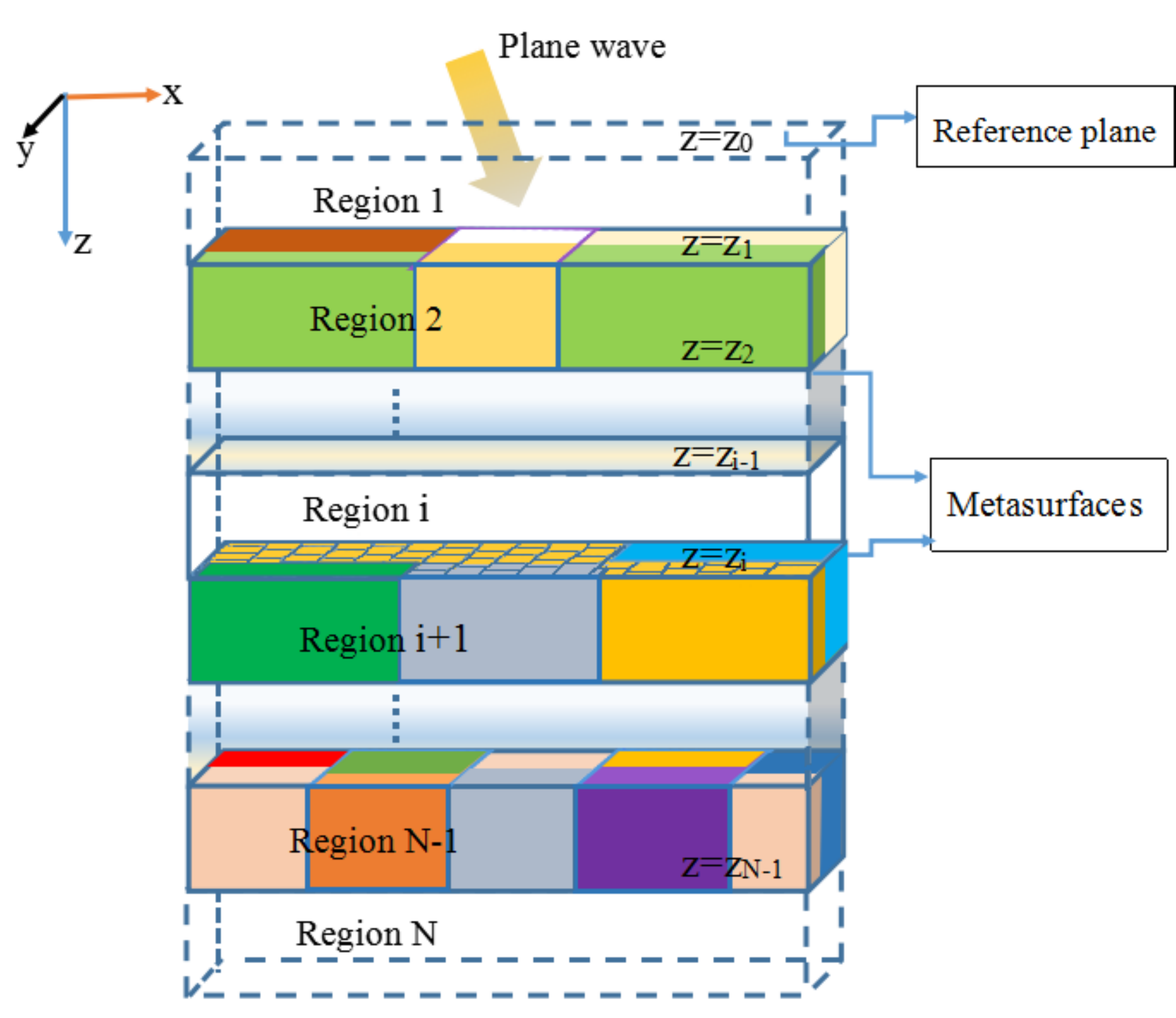}
   }
 \caption{The geometry of a multi-region layered doubly periodic structure. Region $1$ and region $N$ are semi-infinite, so they satisfy the radiation boundary conditions in the $-z$ and $+z$ directions, respectively. The front, back, left and right outer boundaries are periodic, and hence satisfy the Bloch (Floquet) periodic boundary conditions under an incident plane wave. Metasurfaces and other impedance surfaces may appear at any layer interfaces.}
\label{sketch1}
\end{figure}

\IEEEPARstart{T}{hree}-dimensional layered media with doubly periodic structures are ubiquitous in microwave, millimeter wave and optical integrated circuits, electronic packages, and other fields \cite{Wait1970,Chew1985x,Chew1990, QHLiu1990, Johnson2000, Chen2011, Dai2015, Jiao2007, Gan2007, Gan2008, Wang2012}, as illustrated in Figure \ref{sketch1}. Simulating electromagnetic waves interacting with such complex structures is essential for rapid prototyping of devices involving such structures.
More recently, metasurfaces have been widely studied due to their novel electromagnetic and optical properties; they may appear as very thin surfaces at some layer interfaces in Figure \ref{sketch1} \cite{Zhu2013,Zhu2018,Zhu2018x,Vahabzadeh2016,Song2016}.  In optical lithography, the periodic patterns can also be embedded in such a multilayer structure \cite{Niu2017}.

In order to simulate the propagation and action of electromagnetic waves in such a multi-region structure, the three-dimensional (3D) Maxwell's equations need to be solved.  Traditionally, the finite element method (FEM) and finite difference time domain (FDTD) method have been widely applied with success. However, the conventional numerical methods will consume huge computational resources for such complex media, especially when the layer thicknesses are large compared to the wavelength.  Meanwhile, the semi-infinity of region $1$ and region $N$ also increases the difficulty of calculation. To overcome the difficulties mentioned above, there are many excellent algorithms such as the numerical mode-matching (NMM) method \cite{Chew1985x, Chew1990, QHLiu1990, Dai2015}, the semi-analytical spectral element method (SEM) \cite{Chen2011}, the layered finite element (LAFE) method \cite{Jiao2007, Gan2007, Gan2008} and the spectral element boundary integral (SEBI) method \cite{Niu2017}.

In particular, the NMM method has been shown in 2D to be more efficient than the direct use of the conventional numerical methods (e.g., the FDTD, FEM and MOM), because it is a semi-analytical solver to reduce a $d$-dimensional ($d=2,3$) EM field problem into several $(d-1)$-dimensional eigenvalue problems in the horizontal dimensions and an analytical scheme in the vertical dimension. Therefore, the NMM method can significantly reduce the computational costs and achieve more accurate solutions, so that it has been used to model various EM field problems in 2D and 2.5D \cite{QHLiu1990,Dai2015,Wang2012, Chew1984, Chew1991, QHLiu1992, QHLiu1993, Chew1985, Li2005, Hue2007}.

In this paper, a 3-D semi-analytical solver is developed to model the multi-region structure with doubly periodic boundary conditions in the horizontal directions, based on the NMM idea with the mixed spectral element method for the 2-D eigenvalue problem, hence it is called the spectral numerical mode-matching (SNMM) method. In order to obtain these high accuracy physical eigenmodes, the mixed spectral element method (MSEM) is employed to solve the Bloch periodic waveguide eigenvalue problems \cite{Liu2018}. The MSEM is based on the spectral element method (SEM) and Gauss' law, which can remove all the nonphysical eigenmodes and achieve exponential convergence. In general, the SEM has exponential convergence for an appropriate smooth solution because of its use of the basis functions constructed by the high-degree polynomials \cite{Lee2006, Luo2009}. In view of the quality of the SEM, the Gauss-Lobatto-Legendre (GLL) polynomials are used to construct the curl-conforming vector edge-based basis functions and the scalar continuous nodal-based basis functions in the MSEM. Both SEM and MSEM have been widely applied to solve the Maxwell's eigenvalue problems \cite{Lee2006, Luo2009, Luo2009x, Liu2015,Liux2015,Peverini2011}, but MSEM does not generate any spurious modes, so it is used in this work for the 3-D NMM method.

Although both 2-D and 2.5-D NMM methods have been widely reported, the only 3-D NMM method reported so far is for the scalar Poisson's equation \cite{Fan2000}. Recently, the 3-D SNMM method was extended to a two-region problem, with a metasurface separating two half spaces, for Maxwell's equations \cite{Liu2019}.  In this work, we extend this SNMM method to arbitrary multi-region problems with multiple metasurfaces. The 3-D SNMM inherits the excellent qualities of the MSEM and the NMM method so that it can not only efficiently obtain the accurate solution, but also significantly reduce the computational costs, especially when the layer thicknesses are large.

The rest of this paper is organized as follows. In Section II, the process of the SNMM method is first summarized, then the computation for Bloch eigenmodes is presented. The formulations of the excitation vector, local reflection and transmission matrices are briefly summarized from \cite{Liu2019}, while the generalized reflection matrices are derived in detail for multiple regions. In Section III, the accuracy and efficiency of the SNMM method are demonstrated by several numerical examples. In Section IV, a brief conclusion is given.

\section{Formulation}
As shown in Fig. \ref{sketch1}, we will consider the propagation and scattering of electromagnetic waves in the 3-D layered multi-region doubly periodic structure with metasurfaces. This structure consists of $N$ regions, where region $1$ and region $N$ are semi-infinite with the radiation boundary condition. All the interface $z=z_{i}$ ($i=1,2,\ldots, N-1$) between two different regions are parallel to the reference plane $z=z_{0}$. At the same time, metasurfaces may be present at any layer interfaces $z=z_{i}$. We assume that the medium is inhomogeneous and anisotropic with the following forms in region $i$:
\begin{equation}\label{eq:1}
\bar{\bar{\epsilon}}_{r}^{(i)}(x,y)=
\begin{bmatrix} \bar{\bar{\epsilon}}_{rt}^{(i)} &0\\0&\epsilon_{rz}^{(i)}\end{bmatrix},
\bar{\bar{\mu}}_{r}^{(i)}(x,y)=
\begin{bmatrix} \bar{\bar{\mu}}_{rt}^{(i)} &0\\ 0&\mu_{rz}^{(i)}\end{bmatrix}
\end{equation}
where $\bar{\bar{\epsilon}}_{rt}^{(i)}$ and $\bar{\bar{\mu}}_{rt}^{(i)}$ are full $2\times 2$ tensors; $\bar{\bar{\epsilon}}_{r}^{(i)}$ and $\bar{\bar{\mu}}_{r}^{(i)}$ may be lossy (complex), anisotropic, and arbitrarily inhomogeneous in $(x,y)$ but piece-wise constant in $z$.  Within each region (layer) of Figure \ref{sketch1}, the medium is uniform in $z$.  The SNMM method is described in more detail below with four steps: (a) The solution of Bloch (Floquet) Eigenmodes for each region; (b) source excitation vector; (c) the local reflection and transmission matrices; and (d) global (generalized) reflection and transmission matrices.

\subsection{Bloch (Floquet) Eigenmodes}

The first step of the SNMM method is to find the Bloch (Floquet) eigenmodes of the individual region in Figure \ref{sketch1} by assuming that region to be infinitely long in the $z$ direction. This eigenvalue problem can be solved by using the MSEM for the following waveguide problem with the Bloch periodic boundary conditions (BPBCs) (see \cite{Liu2018}):
\begin{subequations}\label{eq:2}
\begin{align}
\nabla_{t}\times\mu_{rz}^{(i)-1}\nabla_{t}\times\textbf{e}_{t}&+\bar{\bar{R}}\bar{\bar{\mu}}_{rt}^{(i)-1}\bar{\bar{R}}\nabla_{t}e_{z}^{new}\nonumber\\
&-k_{0}^{2}\bar{\bar{\epsilon}}_{rt}^{(i)}\textbf{e}_{t}
=k_{z}^{2}\bar{\bar{R}}\bar{\bar{\mu}}_{rt}^{(i)-1}\bar{\bar{R}}\textbf{e}_{t},\label{eq:2a}\\
&\nabla_{t}\cdot(\bar{\bar{\epsilon}}_{rt}^{(i)}\textbf{e}_{t})-\epsilon_{rz}^{(i)}e_{z}^{new}=0,\label{eq:2b}
\end{align}
\end{subequations}
where $\nabla_{t}=\hat{x}\frac{\partial}{\partial x}+\hat{y}\frac{\partial}{\partial y}$, $e_{z}^{new}=jk_{z}e_{z}$, $k_{z}$ is the propagation constant along the $+z$-direction, $k_{0}$ denotes the wave number in vacuum, and the rotation matrix $\bar{\bar{R}}$ is equivalent to the operator $\hat{z}\times$.
The Bloch periodic boundary conditions for the BPBC waveguide problem (\ref{eq:2}) are shown as
\begin{equation}\label{eq:3}
\textbf{e}_{t}(\textbf{r}+\textbf{a})=\textbf{e}_{t}(\textbf{r})e^{-j\textbf{k}_{t}\cdot\textbf{a}},~
e_{z}^{new}(\textbf{r}+\textbf{a})=e_{z}^{new}(\textbf{r})e^{-j\textbf{k}_{t}\cdot\textbf{a}}
\end{equation}
where $\textbf{k}=\textbf{k}_{t}+\hat{z}k_{z}$ is the Bloch wave vector, $\textbf{r}$ is the position vector on the boundary $\partial\Gamma$ of the cross section $\Gamma$ of the BPBC waveguide and $\textbf{a}=\hat a_1 a_1+\hat a_2 a_2$ (both unit vectors $\hat a_1$ and $\hat a_2$ are perpendicular to $\hat z$) denotes the lattice translation vector.

As explained in \cite{Liu2018}, the eigenfunctions $\textbf{e}_{t}$ and $e_{z}^{new}$ are further written as the plane wave forms
\begin{equation}\label{eq:4}
\textbf{e}_{t}(\textbf{k}_{t},\textbf{r})=\textbf{u}(\textbf{k}_{t},\textbf{r})e^{-j\textbf{k}_{t}\cdot\textbf{r}},
e_{z}^{new}(\textbf{k}_{t},\textbf{r})=w(\textbf{k}_{t},\textbf{r})e^{-j\textbf{k}_{t}\cdot\textbf{r}}.
\end{equation}
From (\ref{eq:3}), the following periodic boundary conditions can be obtained by the first corollary of Bloch theorem \cite{Bloch1929}
\begin{equation}\label{eq:5}
\textbf{u}(\textbf{k}_{t},\textbf{r})=\textbf{u}(\textbf{k}_{t},\textbf{r}+\textbf{a}),~
w(\textbf{k}_{t},\textbf{r})=w(\textbf{k}_{t},\textbf{r}+\textbf{a}).
\end{equation}
Substituting (\ref{eq:4}) into (\ref{eq:2}), we can achieve the PBC waveguide problem:
\begin{subequations}\label{eq:6}
\begin{align}
(&\nabla_{t}-j\textbf{k}_{t})\times\mu_{rz}^{(i)-1}(\nabla_{t}-j\textbf{k}_{t})\times\textbf{u}\nonumber\\
&+\bar{\bar{R}}\bar{\bar{\mu}}_{rt}^{(i)-1}\bar{\bar{R}}(\nabla_{t}-j\textbf{k}_{t})w
-k_{0}^{2}\bar{\bar{\epsilon}}_{rt}^{(i)}\textbf{u}=k_{z}^{2}\bar{\bar{R}}\bar{\bar{\mu}}_{rt}^{(i)-1}\bar{\bar{R}}\textbf{u} \label{eq:15a}\\
&(\nabla_{t}-j\textbf{k}_{t})\cdot(\bar{\bar{\epsilon}}_{rt}^{(i)}\textbf{u})-\epsilon_{rz}^{(i)}w=0,\label{eq:15b}
\end{align}
\end{subequations}
Along the way of  \cite{Liu2018}, the eigenpair $(\textbf{u}, w, k_{z})$ can be obtained by using the MSEM. It is easy to see that once the eigenpair $(\textbf{u}, w, k_{z})$ is obtained from (\ref{eq:6}), then the eigenpair $(\textbf{e}_{t}, e_{z}^{new}, k_{z})$ can be also obtained from (\ref{eq:4}). Another advantage of doing this is that the excitation coefficient $f_{\alpha,s}$ can be easily solved by using the eigenfunction $\textbf{u}$ in what follows.

\subsection{Excitation Vector}

Now if a plane wave is incident from region 1 or region $N$, assuming this region is infinitely long along the $z$ direction, one can expand this plane wave in terms of the summation of the above eigenmodes in that region. The collection of the expansion coefficients forms the excitation vector in this region for the incident wave, and they represent the amount of eigenmodes being excited. As detailed in \cite{Liu2019}, the incident transverse electric and magnetic fields $\textbf{E}_{t}^{\textrm{inc}}$ and $\textbf{H}_{t}^{\textrm{inc}}$ are first written as:
\begin{equation}\label{eq:7}
\begin{split}
\textbf{E}_{t}^{\textrm{inc}}
=\bar{\bar{\textbf{F}}}^{t} e^{-j\bar{\bar{K}}_{z}(z-z_{0})}\textbf{F}_{s},
\end{split}
\end{equation}
\begin{equation}\label{eq:8}
\bar{\bar{R}}\textbf{H}_{t}^{\textrm{inc}}=\mathcal{N}\bar{\bar{\textbf{F}}}^{t}e^{-j\bar{\bar{K}}_{z}(z-z_{0})}\textbf{F}_{s},
\end{equation}
where the operator $\mathcal{N}=\frac{1}{k_{z}\omega\mu_{0}}\{\nabla_{t}\times(\mu_{rz}^{-1}\nabla_{t}\times)
-k_{0}^{2}\bar{\bar{\epsilon}}_{rt}\cdot\}$; $\textbf{F}_{s}=\{f_{1,s},f_{2,s},\ldots,f_{m,s}\}^{t}$ is the excitation vector, $\bar{\bar{\textbf{F}}}^{t}=\{\textbf{e}_{1,t},\textbf{e}_{2,t},\ldots,\textbf{e}_{m,t}\}$, and $\bar{\bar{K}}_{z}=\textrm{diag}\{k_{1,z},k_{2,z},\ldots,k_{m,z}\}$.
Moreover, a uniform plane wave is written as
\begin{equation}\label{eq:9}
\textbf{E}^{\textrm{inc}}=\textbf{E}_{0}e^{-j\textbf{k}_{t}\cdot\textbf{r}_{t}}e^{-jk_{z}(z-z_{0})},
\end{equation}
where the wave vector $\textbf{k}(\theta_{k},\phi_{k})=\textbf{k}_{t}+\hat{z}k_{z}$ and the constant vector $\textbf{E}_{0}(\phi_{e})$ can be found from \cite{Liu2019};
$(\theta_{k},\phi_{k})$ are the elevation and azimuthal angles of the propagation direction, $\phi_{e}$ is azimuthal angles of electric field vector.

On the one hand, substituting the left-hand side of (\ref{eq:7}) with the transverse components $\textbf{E}_{0,t}e^{-j\textbf{k}_{t}\cdot\textbf{r}_{t}}e^{-jk_{z}(z-z_{0})}$, taking $z=z_{0}$ and noting that (\ref{eq:4}), we obtain
\begin{equation}\label{eq:10}
\textbf{E}_{0,t}e^{-j\textbf{k}_{t}\cdot\textbf{r}_{t}}=\sum_{\alpha=1}^{m}\textbf{u}_{\alpha}e^{-j\textbf{k}_{t}\cdot\textbf{r}_{t}} f_{\alpha,s}.
\end{equation}
On the other hand, for all $\beta=1,2,\ldots,m$
\begin{equation*}
\bar{\bar{R}}\textbf{h}_{\beta,t}=\mathcal{N}\textbf{e}_{\beta,t}\equiv \tilde{\mathcal{N}}\textbf{u}_{\beta}e^{-j\textbf{k}_{t}\cdot\textbf{r}_{t}}
\end{equation*}
where the operator $\tilde{\mathcal{N}}=\frac{1}{k_{\beta,z}\omega\mu_{0}}(\nabla_{t}-j\textbf{k}_{t})\times(\mu_{rz}^{-1}(\nabla_{t}-j\textbf{k}_{t})\times)-k_{0}^{2}\bar{\bar{\epsilon}}_{rt}\cdot$
Multiplying (\ref{eq:10}) by $\bar{\bar{R}}\textbf{h}_{\beta,t}$ and integrating, we obtain a linear system
\begin{equation}\label{eq:11}
\bar{\bar{Y}}_{0}\textbf{F}_{s}=\textbf{b},
\end{equation}
where $\textbf{b}$ and $\bar{\bar{Y}}_{0}$ consist of the elements, respectively
\begin{equation*}
b_{\beta}=(\tilde{\mathcal{N}}\textbf{u}_{\beta}, \textbf{E}_{0,t}),~~
[\bar{\bar{Y}}_{0}]_{\beta,\alpha}=(\tilde{\mathcal{N}}\textbf{u}_{\beta},\textbf{u}_{\alpha}).
\end{equation*}
Therefore, the excitation vector $\textbf{F}_{s}$ can be obtained by solving the linear system (\ref{eq:11}).

Specifically, for a inhomogeneous lossless or a homogeneous lossy media $\bar{\bar{\epsilon}}_{r}$ and $\bar{\bar{\mu}}_{r}$ in region 1 or region $N$, according to the method of \cite{Chew1989}, the following orthogonality can be proven
\begin{equation}\label{eq:12}
C\delta_{\alpha\beta}=(\textbf{e}_{\alpha,t},\bar{\bar{R}}\textbf{h}_{\beta,t})\equiv (\textbf{u}_{\alpha}, \tilde{\mathcal{N}}\textbf{u}_{\beta}),~ \alpha, \beta=1,2,\ldots
\end{equation}
where $C$ is a constant and $\delta_{\alpha\beta}$ denotes the Kronecker delta. Under the inter product (\ref{eq:12}), by using the Gram-Schmidt orthogonalization, the eigenfunction sequence $\textbf{u}_{\alpha}$ can be normalized orthogonally. As a result, the matrix $\bar{\bar{Y}}_{0}$ is a unit matrix so that it is easy to obtain the excitation vector $\textbf{F}_{s}=\textbf{b}$ from (\ref{eq:11}), which is equation (24) of \cite{Liu2019}.

\subsection{Local Reflection and Transmission Matrices}

The above considers the eigenmodes and excitation vector for an infinitely thick layer. Now consider two adjacent layers (say layers $i$ and $i+1$) in Figure \ref{sketch1} forming two half spaces separated by a metasurface at their interface $z=z_{i}$. Then the waves incident at this interface will be reflected and transmitted due to the eigenmode conversion at the interface. The local reflection and transmission matrices between the adjacent regions can be deduced by using the boundary conditions on the interface $z=z_{i}$. Here we assume that the incident waves impinge on the interface from region $i$ to region $i+1$. Then the fields can be expressed in terms of the reflection and transmission of eigenmodes \cite{QHLiu1990}. As described in \cite{Liu2019}, the total fields $\textbf{E}_{i,t}$ and $\bar{\bar{R}}\textbf{H}_{i,t}$ in region $i$ can be written as:
\begin{equation}\label{eq:13}
\begin{split}
\textbf{E}_{i,t}=\bar{\bar{\textbf{F}}}_{i}^{t}\cdot&[e^{-j\bar{\bar{K}}_{i,z}(z-z_{0})}\\
&+e^{j\bar{\bar{K}}_{i,z}(z-z_{i})}\bar{\bar{R}}_{i,i+1}e^{-j\bar{\bar{K}}_{i,z}(z_{i}-z_{0})}]\textbf{F}_{s},
\end{split}
\end{equation}
\begin{equation}\label{eq:14}
\begin{split}
\bar{\bar{R}}\textbf{H}_{i,t}=\bar{\bar{\textbf{N}}}_{i}^{t}\cdot&[e^{-j\bar{\bar{K}}_{i,z}(z-z_{0})}\\
&-e^{j\bar{\bar{K}}_{i,z}(z-z_{i})}\bar{\bar{R}}_{i,i+1}e^{-j\bar{\bar{K}}_{i,z}(z_{i}-z_{0})}]\textbf{F}_{s},
\end{split}
\end{equation}
where $\bar{\bar{\textbf{N}}}^{t}=\mathcal{N}\bar{\bar{\textbf{F}}}^{t}$.
Similarly, the total fields $\textbf{E}_{i+1,t}$ and $\bar{\bar{R}}\textbf{H}_{i+1,t}$ in region $i+1$ can be written as:
\begin{equation}\label{eq:15}
\begin{split}
\textbf{E}_{i+1,t}=\bar{\bar{\textbf{F}}}_{i+1}^{t}
e^{-j\bar{\bar{K}}_{i+1,z}(z-z_{i})}\bar{\bar{T}}_{i,i+1}e^{-j\bar{\bar{K}}_{i,z}(z_{i}-z_{0})}\textbf{F}_{s},
\end{split}
\end{equation}
\begin{equation}\label{eq:16}
\begin{split}
\bar{\bar{R}}\textbf{H}_{i+1,t}=
\bar{\bar{\textbf{N}}}_{i+1}^{t}e^{-j\bar{\bar{K}}_{i+1,z}(z-z_{i})}\bar{\bar{T}}_{i,i+1}e^{-j\bar{\bar{K}}_{i,z}(z_{i}-z_{0})}\textbf{F}_{s},
\end{split}
\end{equation}
where $\bar{\bar{R}}_{i,i+1}$ and $\bar{\bar{T}}_{i,i+1}$ are the local reflection and transmission matrices, which can be obtained from equation (43) of Appendix B in \cite{Liu2019}.

Conversely, when the incident waves impinge from region $i+1$ to region $i$, the local reflection and transmission matrices $\bar{\bar{R}}_{i+1,i}$ and $\bar{\bar{T}}_{i+1,i}$ can be also defined by reversing the subscripts $i$ and $i+1$ in (\ref{eq:13})-(\ref{eq:16}).
Similarly, the subscripts $i$ and $i+1$ in the boundary condition (20) of \cite{Liu2019} are also alternated, then we can obtain the local reflection and transmission matrices $\bar{\bar{R}}_{i+1,i}$ and $\bar{\bar{T}}_{i+1,i}$ by swapping the role of subscripts $i$ and $i+1$ in the derivation process of Appendix B in \cite{Liu2019}.

\subsection{Global (Generalized) Reflection and Transmission Matrices}

Finally we can discuss the solution to the whole problem in Figure \ref{sketch1}. As shown in Figure 1, we assume that the direction of the wave propagation is always from top to bottom along the $z$-axis. In region $n$ ($n=1,2,\ldots,N$), the global (generalized) reflection matrix $\tilde{\bar{G}}_{n,n+1}$ can be defined to relate the upgoing waves with the downgoing waves, and it can be expressed in terms of the local reflection and transmission matrices \cite{QHLiu1990}.
Consequently, the transverse fields can be expressed in terms of the generalized reflection matrices, propagator matrices, excitation vector and the eigenmodes in region $n$ ($n=1,2,\ldots,N$):
\begin{equation}\label{eq:17}
\begin{split}
\textbf{E}_{n,t}=\bar{\bar{\textbf{F}}}_{n}^{t}\cdot[e^{-j\bar{\bar{K}}_{n,z}(z-z_{n-1})}&+e^{j\bar{\bar{K}}_{n,z}(z-z_{n})}\\
&\cdot \tilde{\bar{G}}_{n,n+1} \bar{\bar{P}}_{n}]\textbf{A}_{n},
\end{split}
\end{equation}
\begin{equation}\label{eq:18}
\begin{split}
\bar{\bar{R}}\textbf{H}_{n,t}=\bar{\bar{\textbf{N}}}_{n}^{t}\cdot[e^{-j\bar{\bar{K}}_{n,z}(z-z_{n-1})}&-e^{j\bar{\bar{K}}_{n,z}(z-z_{n})}\\
&\cdot\tilde{\bar{G}}_{n,n+1}\bar{\bar{P}}_{n}]\textbf{A}_{n},
\end{split}
\end{equation}
where $\textbf{A}_{n}$ is the amplitude of the downgoing wave, and $\bar{\bar{P}}_{n}=e^{-j\bar{\bar{K}}_{n,z}(z_{n}-z_{n-1})}$ is the propagator matrix inside the $n$-th layer from $z_{n-1}$ and $z_n$.

We next derive the generalized reflection matrix $\tilde{\bar{G}}_{n,n+1}$, the amplitude $\textbf{A}_{n}$ and the global transmission matrix $\tilde{\bar{T}}_{1,N}$ by using the physical interpretation of the local reflection and transmission matrices between two adjacent regions. From (\ref{eq:17}), at $z=z_{n}$, the upgoing waves in region $n$ are the result of local reflection of the downgoing waves in region $n$ plus the local transmission of the upgoing waves in region $n+1$, thus
\begin{equation}\label{eq:19}
\begin{split}
\tilde{\bar{G}}_{n,n+1}\bar{\bar{P}}_{n}\textbf{A}_{n}
&=\bar{\bar{R}}_{n,n+1}\bar{\bar{P}}_{n}\textbf{A}_{n}\\
&+\bar{\bar{T}}_{n+1,n}\bar{\bar{P}}_{n+1}\tilde{\bar{G}}_{n+1,n+2}
\bar{\bar{P}}_{n+1}\textbf{A}_{n+1},
\end{split}
\end{equation}
Similarly, at $z=z_{n-1}$, the downgoing waves in region $n$ are the result of the local transmission of the downgoing waves in region $n-1$ plus the local reflection of the upgoing waves in region $n$, thus
\begin{equation}\label{eq:20}
\textbf{A}_{n}=\bar{\bar{T}}_{n-1,n}\bar{\bar{P}}_{n-1}\textbf{A}_{n-1}+
\bar{\bar{R}}_{n,n-1}\bar{\bar{P}}_{n}\tilde{\bar{G}}_{n,n+1}\bar{\bar{P}}_{n}\textbf{A}_{n}.
\end{equation}
From (\ref{eq:20}), we can deduce
\begin{equation*}
[\bar{\bar{I}}-\bar{\bar{R}}_{n,n-1}\bar{\bar{P}}_{n}\tilde{\bar{G}}_{n,n+1}\bar{\bar{P}}_{n}]\textbf{A}_{n}
=\bar{\bar{T}}_{n-1,n}\bar{\bar{P}}_{n-1}\textbf{A}_{n-1},
\end{equation*}
which leads to the following recursive relation:
\begin{equation}\label{eq:21}
\begin{split}
\textbf{A}_{n}=\bar{\bar{M}}_{n}^{-1}\bar{\bar{T}}_{n-1,n}
\bar{\bar{P}}_{n-1}\textbf{A}_{n-1}\equiv \tilde{\bar{T}}_{n-1,n}\textbf{A}_{n-1},
\end{split}
\end{equation}
where $\bar{\bar{M}}_{n}=\bar{\bar{I}}-\bar{\bar{R}}_{n,n-1}\bar{\bar{P}}_{n}\tilde{\bar{G}}_{n,n+1}\bar{\bar{P}}_{n}$, and similar to the global transmission coefficient defined in a 1D planar layered medium formulation shown in \cite{QHLiu_EM571},
\begin{equation*}
\tilde{\bar{T}}_{n-1,n}=\bar{\bar{M}}_{n}^{-1}\bar{\bar{T}}_{n-1,n}\bar{\bar{P}}_{n-1}
\end{equation*}
is called the global transmission matrix from region $n-1$ to region $n$. Therefor,
for all $n=1,2,\ldots,N$, the amplitude $\textbf{A}_{n}$
can be derived from (\ref{eq:21}) by using the initial condition $\textbf{A}_{1}=\textbf{F}_{s}$ which can be obtained from (\ref{eq:13}) or (\ref{eq:14}) when $i=1$.

On the other hand, $\textbf{A}_{n+1}$ can be first obtained from the recursive relation (\ref{eq:21}), and then inserting it into (\ref{eq:19}) gives
\begin{equation*}
\begin{split}
\tilde{\bar{G}}_{n,n+1}\bar{\bar{P}}_{n}\textbf{A}_{n}=\bar{\bar{R}}_{n,n+1}\bar{\bar{P}}_{n}\textbf{A}_{n}
&+\bar{\bar{T}}_{n+1,n}\bar{\bar{P}}_{n+1}\tilde{\bar{G}}_{n+1,n+2}\\
&\cdot \bar{\bar{P}}_{n+1}\bar{\bar{M}}_{n+1}^{-1}\bar{\bar{T}}_{n,n+1}
\bar{\bar{P}}_{n}\textbf{A}_{n},
\end{split}
\end{equation*}
which leads to the recursive relation:
\begin{equation}\label{eq:22}
\begin{split}
\tilde{\bar{G}}_{n,n+1}=\bar{\bar{R}}_{n,n+1}&+\bar{\bar{T}}_{n+1,n}\bar{\bar{P}}_{n+1}\\
&\cdot\tilde{\bar{G}}_{n+1,n+2}\bar{\bar{P}}_{n+1}\bar{\bar{M}}_{n+1}^{-1}\bar{\bar{T}}_{n,n+1}.
\end{split}
\end{equation}
It is clear that when the local reflection and transmission matrices $\bar{\bar{R}}_{i,i+1}$, $\bar{\bar{R}}_{i+1,i}$, $\bar{\bar{T}}_{i,i+1}$ and $\bar{\bar{T}}_{i+1,i}$ are known, then the generalized reflection matrix $\tilde{\bar{G}}_{n,n+1}$ can be derived by using the initial condition $\tilde{\bar{G}}_{N,N+1}=\bar{\bar{0}}$ to the above (\ref{eq:22}), where $\bar{\bar{0}}$ denotes a zero matrix.

From the recursive relation (\ref{eq:21}), we can obtain $\textbf{A}_{N}$ in region $N$
\begin{equation}\label{eq:23}
\textbf{A}_{N}=[\coprod_{\ell=N}^{2}\bar{\bar{M}}_{\ell}^{-1}\bar{\bar{T}}_{\ell-1,\ell}\bar{\bar{P}}_{\ell-1}]\textbf{A}_{1}
\equiv \tilde{\bar{T}}_{1,N}\textbf{A}_{1}
\end{equation}
where again similar to \cite{QHLiu_EM571} for a 1D layered medium,
\begin{equation*}
\tilde{\bar{T}}_{1,N}=\coprod_{\ell=N}^{2}\bar{\bar{M}}_{\ell}^{-1}\bar{\bar{T}}_{\ell-1,\ell}\bar{\bar{P}}_{\ell-1}=\coprod_{\ell=N}^{2} \tilde{\bar{T}}_{\ell-1,\ell}
\end{equation*}
is called the global transmission matrix between region $1$ and region $N$; the symbol '$\coprod$' denotes the inverted order continuous product.

In addition, the absorbance is an important parameter for exploring the characteristics of metasurfaces. As defined in \cite{Liu2019}, by a minor adjustment,
the absorbance can be defined by
\begin{equation}\label{eq:24}
\mathcal{A}_{s}=\frac{|P_{1}|}{P_{\textrm{inc}}}
-\frac{|P_{N}|}{P_{\textrm{inc}}},
\end{equation}
where $P_{\textrm{inc}}$ is the incident power; $P_{1}$ and $P_{N}$ are powers in region $1$ and region $N$, respectively, that is
\begin{equation*}
P_{1}=-\frac{1}{2}\textrm{Re}(\textbf{A}_{1}^{*}\bar{\bar{P}}_{1,+}^{\dag}(\bar{\bar{\textbf{F}}}_{1}^{*},\bar{\bar{\textbf{N}}}_{1}^{t})\bar{\bar{P}}_{1,-}\cdot\textbf{A}_{1}),
\end{equation*}
\begin{equation*}
P_{N}=-\frac{1}{2}\textrm{Re}(\textbf{A}_{N}^{*}\bar{\bar{P}}_{N,+}^{\dag}(\bar{\bar{\textbf{F}}}_{N}^{*},\bar{\bar{\textbf{N}}}_{N}^{t})\bar{\bar{P}}_{N,-}\cdot\textbf{A}_{N}).
\end{equation*}
Here $\bar{\bar{P}}_{n,\pm}=[e^{-j\bar{\bar{K}}_{n,z}(z-z_{n-1})}\pm e^{j\bar{\bar{K}}_{n,z}(z-z_{n})}\tilde{\bar{G}}_{n,n+1}\bar{\bar{P}}_{n}]$, the symbols '*' and '$\dagger$' denote the complex conjugate of the vector and the conjugate transpose of the complex matrix, respectively.
As the discussion of (\ref{eq:12}), when the medium of the $n$-th layer $\bar{\bar{\epsilon}}_{r}^{(n)}$ and $\bar{\bar{\mu}}_{r}^{(n)}$ is inhomogeneous and lossless or homogeneous and lossy, $(\bar{\bar{\textbf{F}}}_{n}^{*},\bar{\bar{\textbf{N}}}_{n}^{t})$ can be a unit matrix due to the inter product (\ref{eq:12}) and the Gram-Schmidt orthogonalization.

\section{Numerical Examples}

In this section, we first present two examples for simulating metasurfaces by the SNMM method, and validate these results with the commercial FEM solver COMSOL. We will compare the CPU time, the number of degrees of freedom (DOF) and the accuracy for our method and COMSOL to show the high accuracy and efficiency of the SNMM method. Second, in order to verify that our method is also efficient for large-scale problems, an example of the extreme ultraviolet (EUV) lithography is simulated. Finally, to compare our results with those from COMSOL, a simplified lithography model is also simulated. The SNMM method is implemented by using Matlab R2014a on a ThinkPad T450 PC. In COMSOL, the perfectly matched layer (PML) absorbing boundary condition is used to truncate the semi-infinite region $1$ and region $N$, and the surface current density is employed to simulate the boundary conditions at $z=z_{i}$ for the metasurface. As shown in Figure \ref{sketch1}, a 3-D model will be directly simulated for the metasurface and the simplified lithography, which leads to a large amount of computation in the traditional FEM. Therefore, we run COMSOL on a server.

For convenience, we introduce the following notations:

 1) $\mathcal{A}_{p}$ denotes the absorbance obtained by the analytical method ($p=a$), the SNMM method ($p=s$) and COMSOL ($p=c$), respectively.

 2) $\textbf{E}_{p,t}$ denotes the transverse electric field obtained by using above methods.

 3) $R\mathcal{A}_{pq}=|\mathcal{A}_{p}-\mathcal{A}_{q}|/|\mathcal{A}_{q}|$ denotes the relative error of absorbance, where $p,q=a,s,c$.

 4) $R\textbf{E}_{pq}=\|\textbf{E}_{p,t}-\textbf{E}_{q,t}\|_{2}/\|\textbf{E}_{q,t}\|_{2}$ denotes the relative error of the transverse electric field, where $p,q=a,s,c$.

\subsection{Multilayer Graphene Surfaces}

\begin{figure}[!t]
  \centering
  {
   \includegraphics[width=0.8\columnwidth,draft=false]{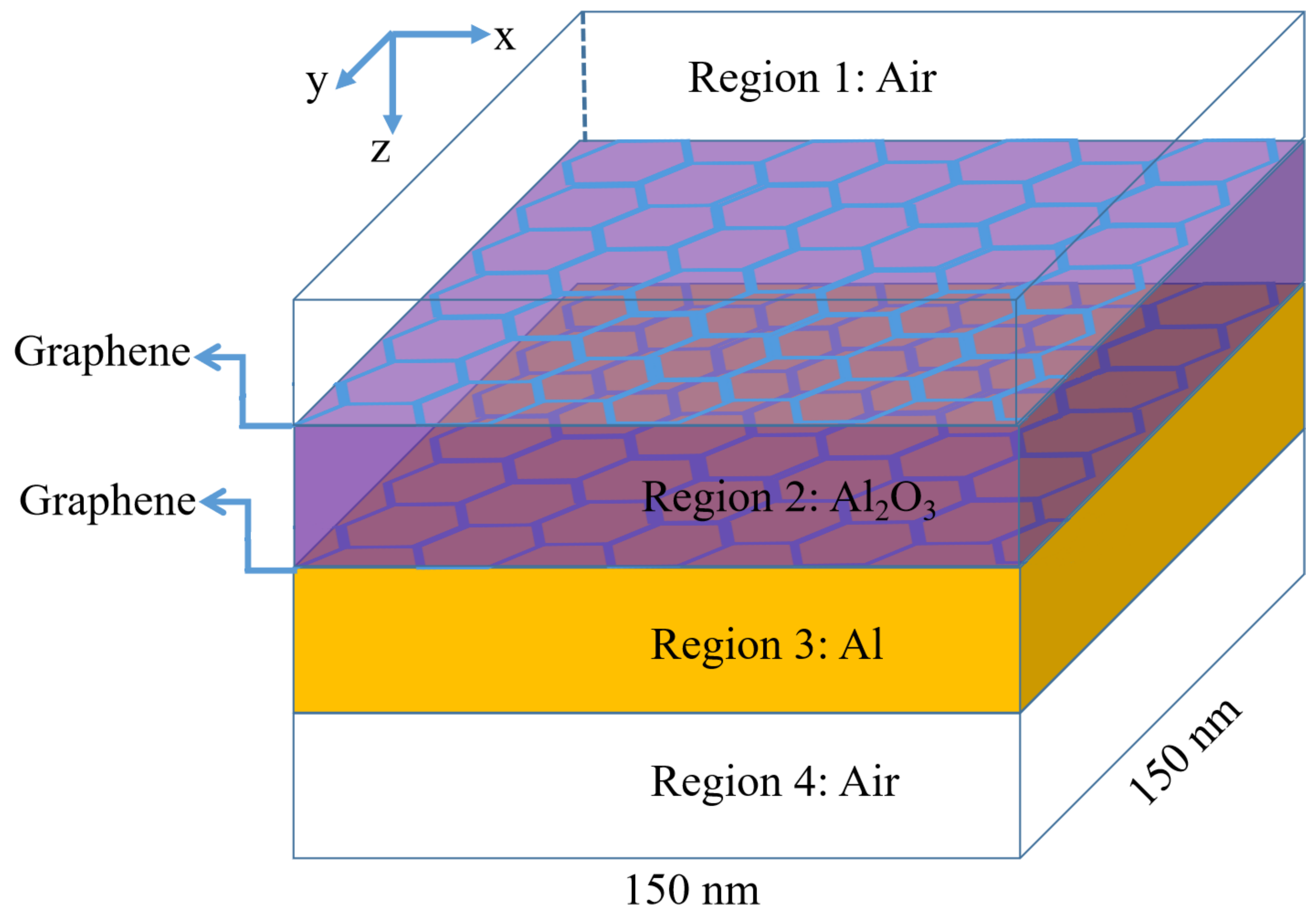}
   }
 \caption{Sketch of the air/graphene/dielectric/metal/air (AGDMA) structure for two graphene surfaces at interfaces of a four-layer medium. Layers 1 and 4 are air, while layer 2 is a 40 nm thick alumina ($\textrm{Al}_{2}\textrm{O}_{3}$) and layer 3 is a 50 nm thick aluminum (Al).}
\label{sketch2}
\end{figure}
The graphene is an important 2D material, and its absorbance is an important parameter for exploring its properties. For a monolayer graphene, its absorbance is about 2.3 $\%$ \cite{Zhu2013, Nair2008}. In order to enhance the absorbance of the graphene, an alternative method is to inlay graphene into the layered structure filled with the dielectric and metal \cite{Zhu2018}. As shown in Fig. \ref{sketch2}, the thickness of the monolayer graphene is much smaller than the operating wavelength, so that it can be assumed as a 2D conductive surface. If the operating wavelength is $\lambda_{0}=314$ nm and the thickness of the graphene is 0.5 nm, then the surface conductivity of graphene can be obtained by the Kubo formula \cite{Francescato2013}, when the temperature $T=300 \textrm{K}$, chemical potential $\mu_{c}=0.33$ eV and charged particle scattering rate $\gamma=0.11\times10^{-3}$ eV. That is, $\tilde{\bar{\sigma}}_{es}=(6.0536\times10^{-5} +j5.8913\times10^{-8})$ S. We assume that a plane wave illuminates this layered structure vertically in the $\hat{z}$ direction and its polarization is $\textbf{E}^{\textrm{inc}}\parallel \hat{y}$, i.e., the elevation and azimuthal angles are set as $(\theta_{k}, \phi_{k}, \phi_{e})=(0,0,\pi/2)$. The 3rd order SNMM method with 2 modes and the 3rd order FEM in COMSOL are used to simulate this example, respectively.

The 40 nm thick alumina ($\textrm{Al}_{2}\textrm{O}_{3}$) and the 50 nm thick aluminum (Al) slabs are selected for the dielectric layer and metal layer, respectively. The refractive index of the alumina (Malitson-e) and the aluminum (Rakic) are $n_{\textrm{Al}_{2}\textrm{O}_{3}}=1.799674$ and $n_{\textrm{Al}}=0.271626-j3.651886$ at $\lambda_{0}=314$ nm, respectively. We will consider four structures filled with different media (i.e., different variations of Figure \ref{sketch2}) to compare the absorbance of graphene, i.e., air/graphene/air (AGA), air/graphene/dielectric/metal/air (AGDMA), air/dielectric/graphene/metal/air (ADGMA) and air/graphene/dielectric/graphene/metal/air (AGDGMA).
\begin{table}[!t]
\renewcommand{\arraystretch}{1.3}
\caption{Absorbance for four different structures}
\centering
\begin{tabular}{cccc}
\hline
Structure type & SNMM & Analytical & COMSOL\\
\hline
AGA    & 0.022294 & 0.022294 & 0.022273\\

AGDMA  & 0.212999 & 0.212999 & 0.213070\\

ADGMA  & 0.170901 & 0.170901 & 0.171036\\

AGDGMA & 0.224288 & 0.224288 & 0.224352\\
\hline
\end{tabular}
\label{example11}
\end{table}

From Table \ref{example11}, first, we can see that the results obtained by the SNMM method agree with the analytical solutions, and they can well match with the results from COMSOL. Second, the results of the AGA structure (monolayer graphene) match well with the experimental result for a 2.3 $\%$ absorptance. Finally, the degree of improvement in the absorbance of the above structures can be arranged as: ADGMA $<$ AGDMA $<$ AGDGMA, where the absorbance of the AGDGMA structure is  about 10 times higher than the AGA structure.

\begin{figure}[!t]
  \centering
  {
   \includegraphics[width=1.0\columnwidth,draft=false]{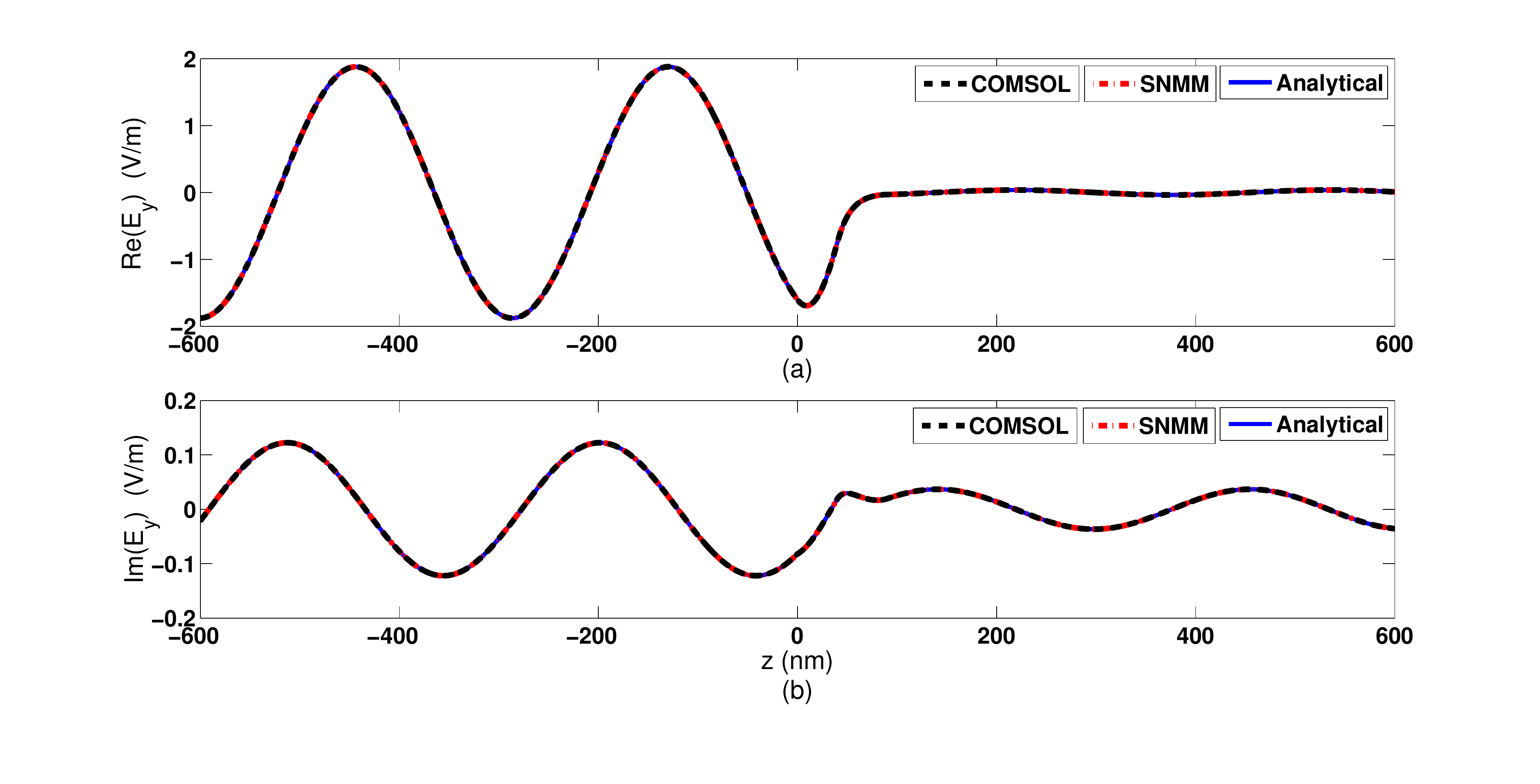}
   }
 \caption{Electric field component $E_{y}$ along the $z$-axis at $(x,y)=(0,0)$ for the AGDGMA structure. (a) Real part of $E_{y}$. (b) Imaginary part of $E_{y}$.}
\label{tu11}
\end{figure}

For simplicity, the results from the most complex AGDGMA structure are compared in detail here. Fig. \ref{tu11} shows that the component $E_{y}$ obtained by the SNMM method, analytical solution and COMSOL are well matched. The relative errors of the transverse electric field are $R\textbf{E}_{sa}=1.04\times10^{-14}$ and
 $R\textbf{E}_{ca}=1.60\times10^{-4}$. The relative errors of the absorbance are $R\mathcal{A}_{sa}= 9.03\times10^{-15}$ and $R\mathcal{A}_{ca}=2.85\times10^{-4}$. The CPU time of COMSOL and the SNMM method are 52 s and 6.45 s, respectively, i.e., the SNMM method is about 8 times faster than COMSOL. COMSOL requires 320919 (or 713 times) more DoF than the SNMM method of 450. From the above discussions, we can see that our SNMM method is highly accurate and efficient for the multilayer graphene surfaces.

\subsection{Gradient Metasurface at Optical/Microwave Frequencies}

The gradient metasurface is first proposed by Capasso's team to generalize Snell's law in 2011 \cite{Yu2011}. It can flexibly and effectively control the phase, polarization states and propagation modes of electromagnetic waves, thus realizing novel physical effects such as anomalous reflection/refraction, surface waves, and so on. Sun \emph{et al.} in \cite{Sunx2012} designed a specific gradient metasurface to demonstrate these properties. Here we follow the work of \cite{Sunx2012} and consider such a metamaterial (MM) slab with a non-uniform $\epsilon_{M}(x)$ and a constant $\mu_{M}(x)$:
\begin{equation}\label{eq:25}
\epsilon_{M}(x)=1+\frac{\xi x}{2k_{0}d},~\mu_{M}(x)=1,\quad 0\le x\le a_{1}
\end{equation}
where $\xi$ is the component of phase gradient parallel to the incident plane, $a_{1}$ is the unit cell length in the $x$ direction, and $d$ denotes the thickness of the MM slab. From (\ref{eq:25}) and the identity $\epsilon_{M}(x)=-j\tilde{\bar{\sigma}}_{es}/\omega\epsilon_{0}d$, we can derive the surface conductivity $\tilde{\bar{\sigma}}_{es}(x)=j(k_{0}d+0.5\xi x)/(120\pi)$ for this MM slab.
\begin{figure}[!t]
  \centering
  {
   \includegraphics[width=0.8\columnwidth,draft=false]{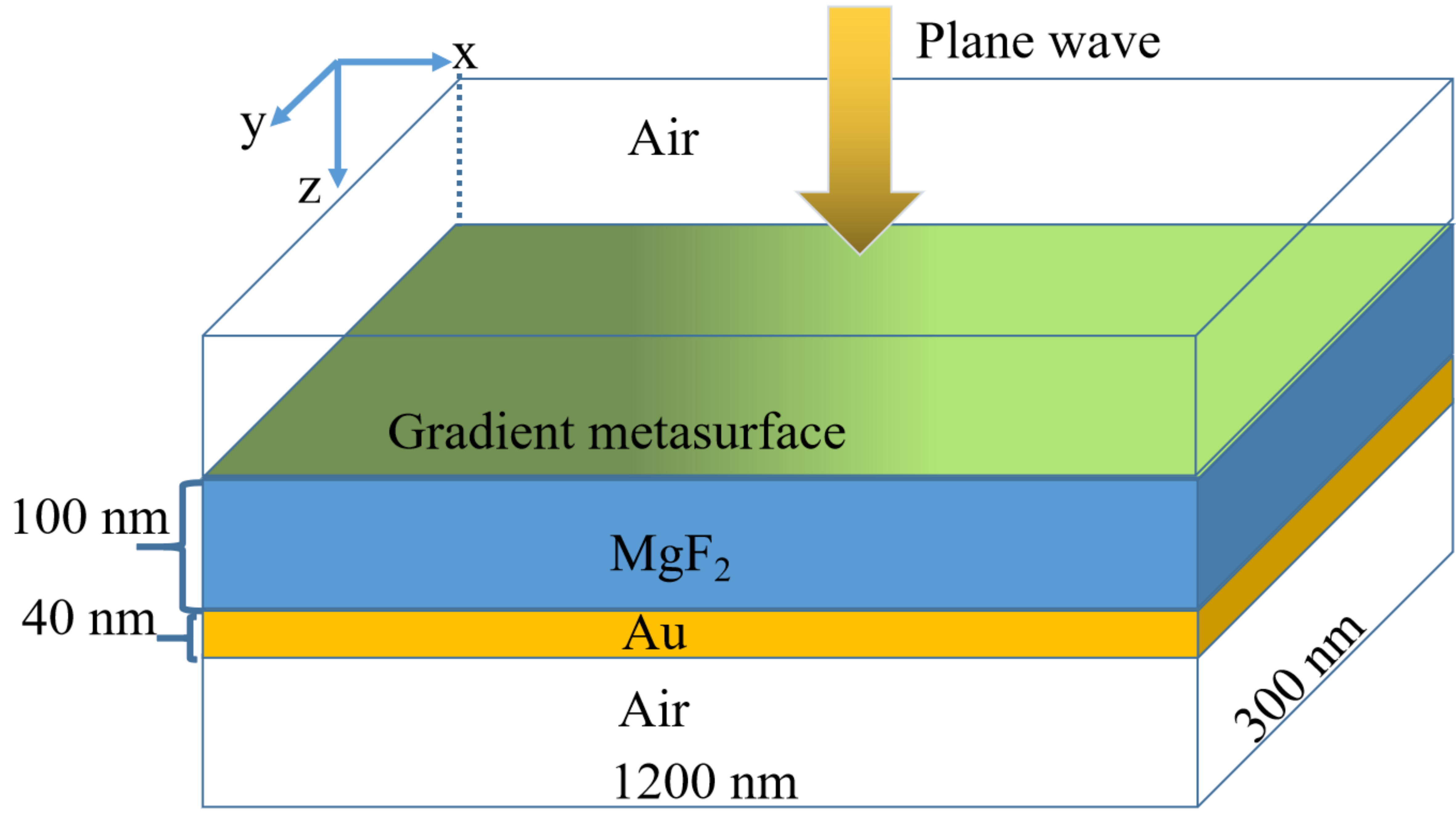}
   }
 \caption{Schematic drawing for the gradient metasurface in the optical frequency range.  For the microwave frequency range studied below, Au is replaced by Titania ($\epsilon_{r}^{(3)}=85$), and all dimensions are multiplied by $10^5$.}
\label{sketch3}
\end{figure}

As shown in Fig. \ref{sketch3}, in order to explore the characteristics of this MM slab, the electric field is solved in the 3D layered structure with $4$ regions. The MM slab has dimensions $1200~\textrm{nm}\times300~\textrm{nm}$ and is adhered on the 100 nm thick $\textrm{MgF}_{2}$ ($\epsilon_{r}^{(2)}=1.892$) substrate. A 40 nm thick Au mirror is placed under the $\textrm{MgF}_{2}$ substrate, and its relative permittivity is $\epsilon_{r}^{(3)}=-27.8343-j2.2176$ when the working wavelength is $850~\textrm{nm}$. The thickness of this MM slab is $d=30~\textrm{nm}$ and $\xi\approx 0.7083 k_{0}$. Here the light is normally incident and polarized with $\textbf{E}^{\textrm{inc}}\parallel \hat{y}$, i.e., the elevation and azimuthal angles are set as $(\theta_{k}, \phi_{k}, \phi_{e})=(0,0,\pi/2)$. The 3rd order SNMM method with 200 Bloch eigenmodes is employed to this example.

\begin{figure}[!t]
  \centering
  {
   \includegraphics[width=0.75\columnwidth,draft=false]{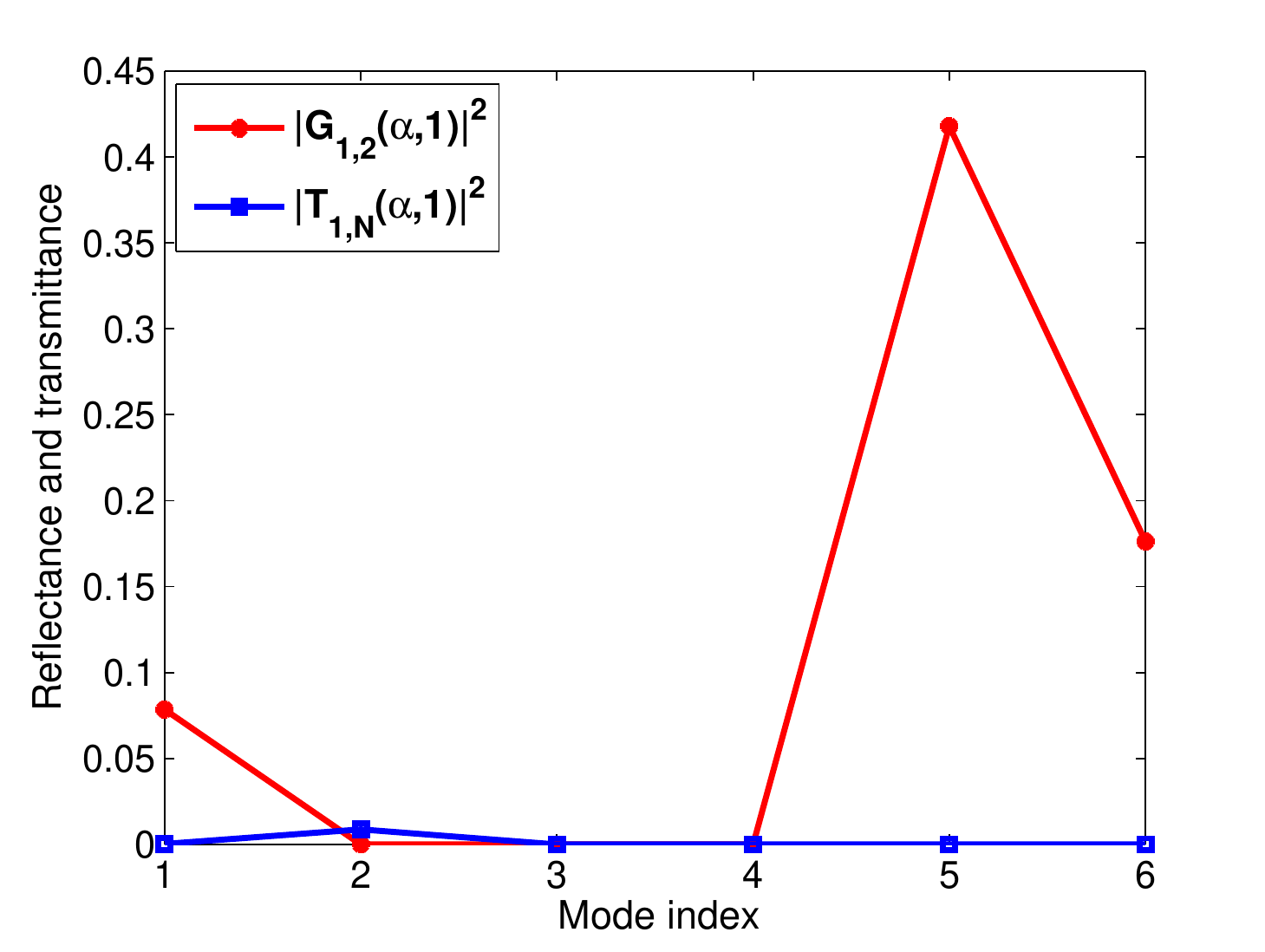}
   }
 \caption{Optical frequency case in Figure \ref{sketch3}: the reflectance and transmittance of the first six propagation modes when the incident wave is the first mode. $\textbf{G}_{1,2}(\alpha,1)$ and $\textbf{T}_{1,N}(\alpha,1)$ are the $\alpha$-th element of the first column of the generalized reflection matrix $\tilde{\bar{G}}_{1,2}$ and the global transmission matrix $\tilde{\bar{T}}_{1,N}$, respectively.}
\label{tu21}
\end{figure}

\begin{figure}[!t]
  \centering
  {
   \includegraphics[width=0.85\columnwidth,draft=false]{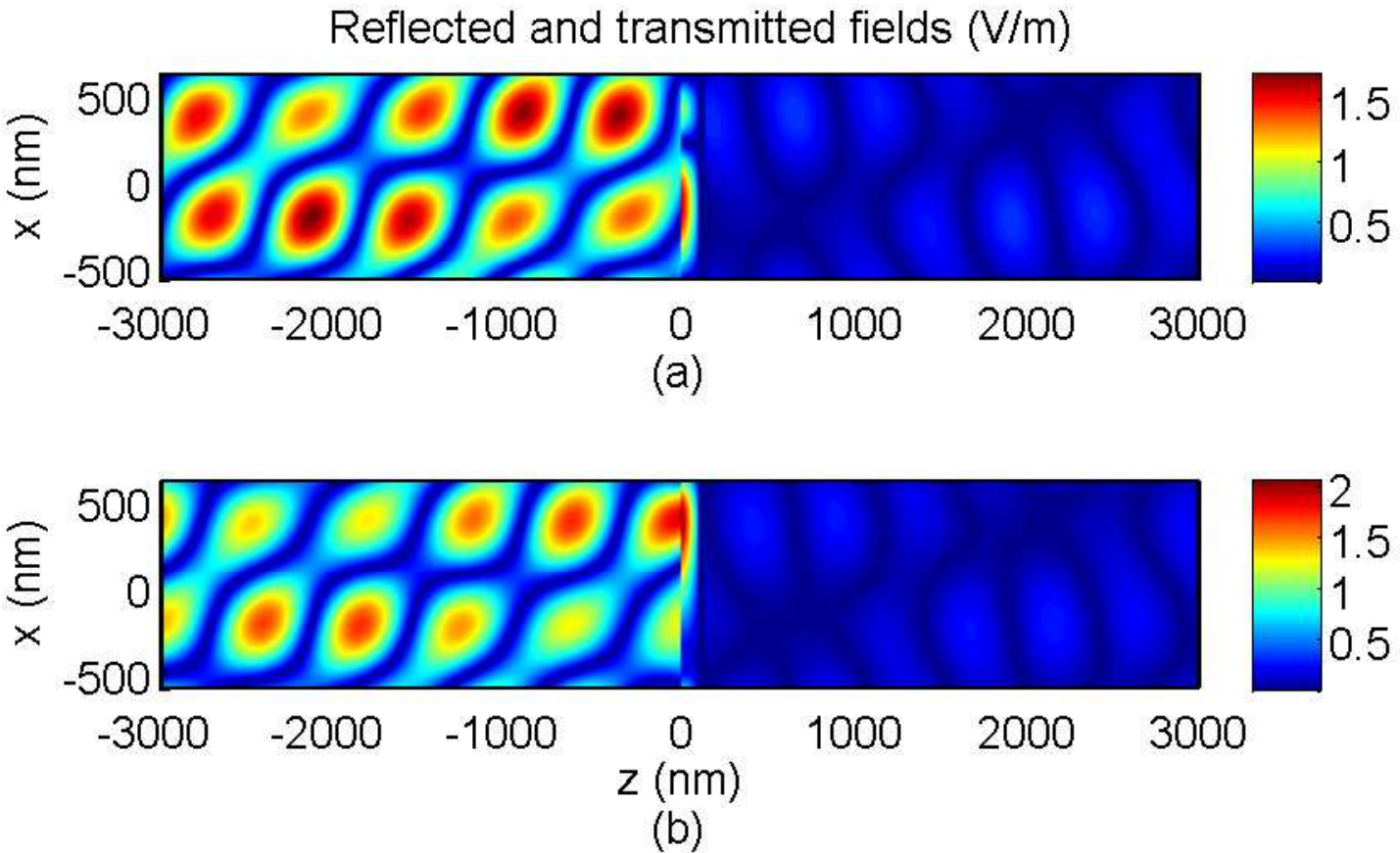}
   }
 \caption{Optical frequency case  in Figure \ref{sketch3}: the field distribution of $E_{y}$ in $xOz$-plane view with $y=0$ nm. (a) Real part of $E_{y}$. (b) Imaginary part of $E_{y}$. }
\label{tu22}
\end{figure}

From the generalized Snell's law of reflection, the angle of reflection $\theta_{r}$ can be written as
\begin{equation}\label{eq:26}
\theta_{r}=\arcsin(\sin\theta_{k}+\frac{\xi}{k_{0}}).
\end{equation}
Thus, $\theta_{r}=\arcsin(0.7083)\approx 45.0968~\textrm{deg}$, when $\theta_{k}=0$ and $\xi\approx 0.7083 k_{0}$. There are six propagation modes for this example, and they have the angle of incidence
\begin{equation*}
\theta_{\alpha,k}=\arccos(\frac{k_{\alpha,z}^{N_{0},h}}{|\textbf{k}|}), ~\alpha=1,2,\ldots, 6.
\end{equation*}
From Table \ref{example21}  for Bloch modes 1-6, we can observe that there are three pairs of degenerate eigenmodes and there exist indeed the modes with the angle of incidence $45.1^{0}$. Furthermore, the polarization of the first eigenmode of each pair is made aligned with the $\hat{y}$ direction, and the second one perpendicular to the $\hat{y}$ direction. After such an operation, $\textbf{F}_{s}$ indicates that the first eigenmode is excited by the incident plane wave. Fig. \ref{tu21} shows that the first, fifth and sixth modes are reflected by this MM slab, and the second mode is refracted to the fourth region, which will lead to anomalous reflection and refraction. From Fig. \ref{tu22}, we can see that the reflected wave appears at about $45^{0}$, which perfectly matches with the theoretical prediction $\theta_{r}\approx 45.0968~\textrm{deg}$. Meanwhile, it is easy to see that there exist indeed anomalous refraction.

\begin{figure}[!t]
  \centering
  {
   \includegraphics[width=1.0\columnwidth,draft=false]{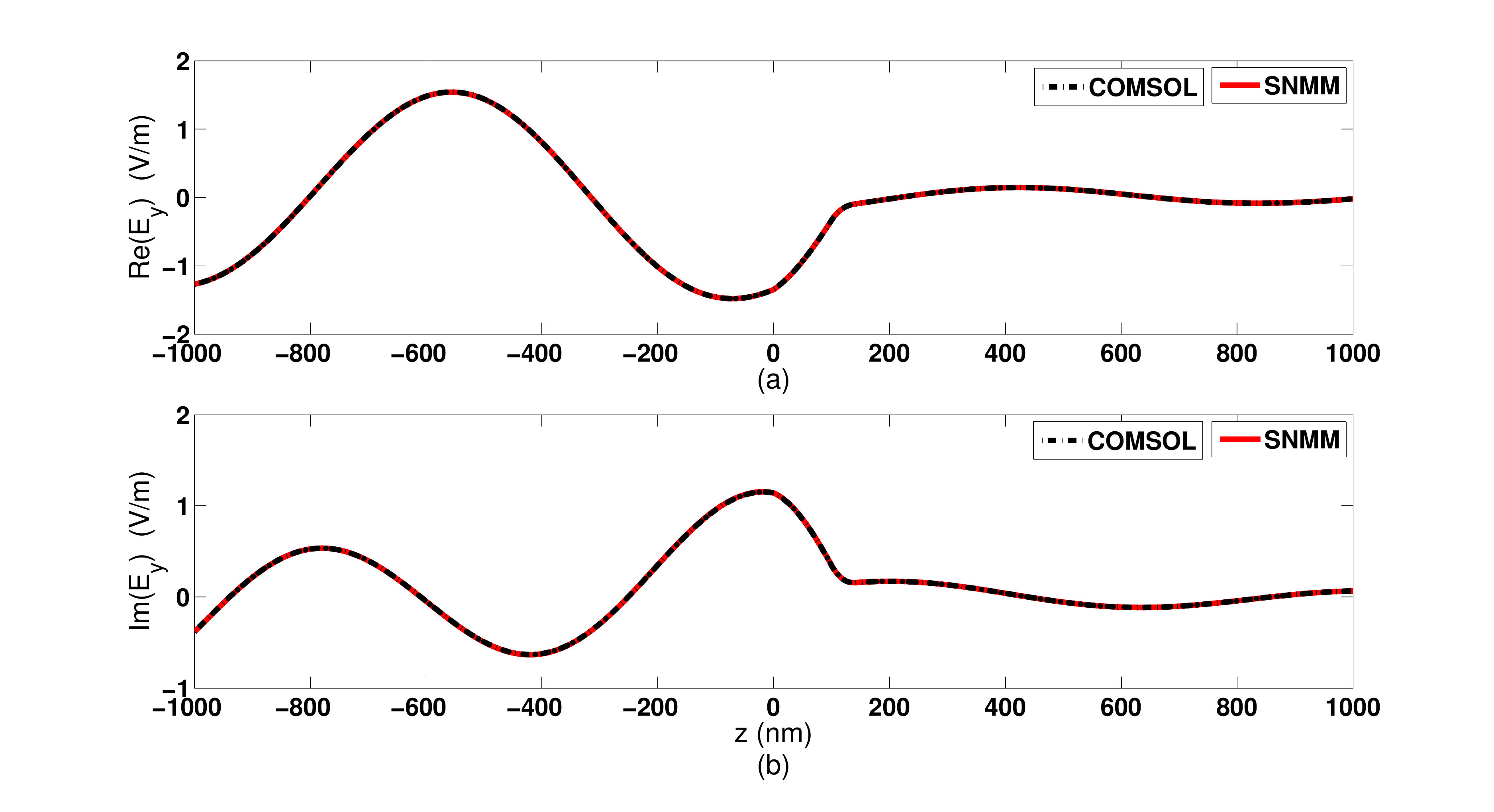}
   }
 \caption{Optical frequency case  in Figure \ref{sketch3}: the electric field component $E_{y}$ along the $z$-axis at $(x,y)=(0,0)$ for the gradient metasurface. (a) Real part of $E_{y}$. (b) Imaginary part of $E_{y}$.}
\label{tu23}
\end{figure}

 The numerical results are listed in Table \ref{example22}. It is easy to see that the absorbance $\mathcal{A}_{s}$ agrees with $\mathcal{A}_{c}$ obtained by COMSOL. The relative error between them is $R\mathcal{A}_{sc}=2.97\times10^{-3}$. The computational speed of the SNMM method is 28.36 times faster than COMSOL. COMSOL requires 7555.8 times the DoF in the SNMM method. Fig. \ref{tu23} shows that the component $E_{y}$ obtained by COMSOL and the SNMM agree well with the relative error
 $R\textbf{E}_{cs}=1.75\times10^{-3}$.
\begin{table}[h]
\renewcommand{\arraystretch}{1.3}
\caption{Parameters of the Propagation Modes for the Gradient Metasurface at Optical Frequency}
\centering
\begin{tabular}{cccc}
\hline
Mode Index & $k_{\alpha,z}^{N,h}$ &$\theta_{\alpha,k} (\textrm{deg})$&$\textbf{F}_{s}$\\
\hline
1 & 7391982.714329 & 0.00000121  & -4.43518963j\\

2 & 7391982.714329 & 0.00000191  & 0.00000000\\

3 & 5217850.164271 & 45.09934657  & 0.00000000\\

4 & 5217850.164271 & 45.09934657  & 0.00000000\\

5 & 5217805.624285 & 45.09983394  & 0.00000000\\

6 & 5217805.624285 & 45.09983394  & 0.00000000\\

\hline
\end{tabular}
\label{example21}
\end{table}

\begin{table}[!t]
\renewcommand{\arraystretch}{1.3}
\caption{Numerical results for gradient metasurface at Optical Frequency}
\centering
\begin{tabular}{cccc}
\hline
Solver & Absorbance & CPU time (s) & DOF\\
\hline
COMSOL & 0.049247 & 536.666667 & 3264117\\

SNMM   & 0.049394 & 18.923958 & 432\\
\hline
\end{tabular}
\label{example22}
\end{table}

\begin{figure}[!t]
  \centering
  {
   \includegraphics[width=0.85\columnwidth,draft=false]{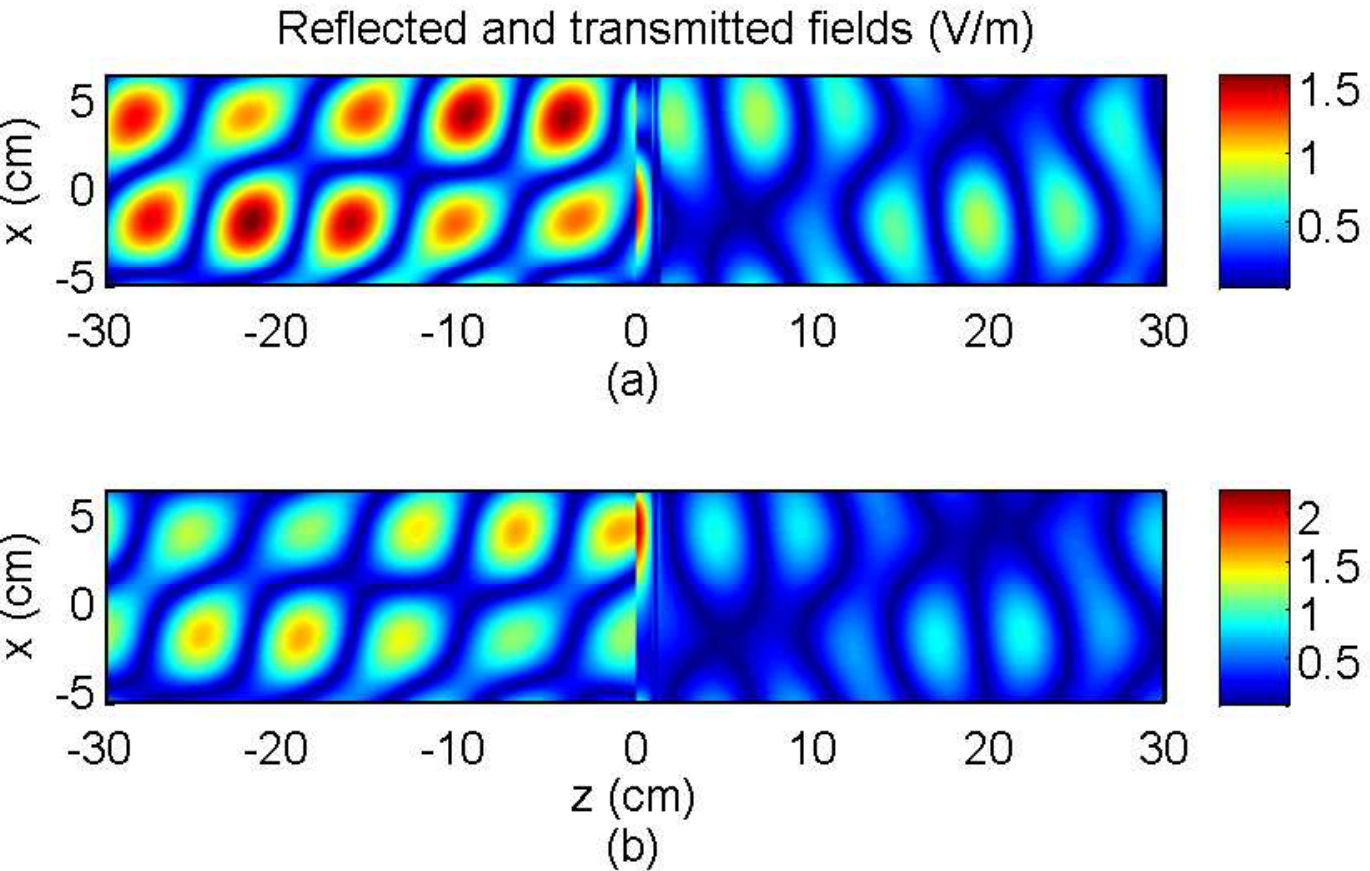}
   }
 \caption{Microwave frequency case in Figure \ref{sketch3}: the field distribution of $E_{y}$ in $xOz$-plane view with $y=0$ nm. (a) Real part of $E_{y}$. (b) Imaginary part of $E_{y}$. }
\label{tu24}
\end{figure}

To show one application in the microwave frequency range where $\lambda_0=8.5$ cm, all physical dimensions of  Figure \ref{sketch3} are multiplied by a factor of $10^5$; and the gold layer is replaced by titania ($\epsilon_r^{(3)}=85$). When the normally incident plane wave is polarized in the $\hat{y}$ direction (mode 1), the
reflectance and transmittance are shown in Table \ref{example23}. Clearly, one also observe that in addition to the small normal reflection of the first mode ($y$ polarization) at about 8\% reflectance, the anomalous reflections into the fifth mode (horizontal polarization) and sixth mode (vertical polarization) at $45^{0}$ reflection angle have reflectance of 29.7\% and $16.0\%$ respectively, which are caused by the MM slab.  Similarly, the anomalous refraction to the fourth region occurs for the second mode ($x$ polarization) at the normal direction, see from Fig. \ref{tu24}. In this case of microwave frequency, the absorbance $\mathcal{A}_{p}$ ($p=s,c$) is approximately zero as all materials are lossless; the relative error $R\textbf{E}_{cs}=2.60\times10^{-3}$. Similarly, the SNMM method is 41.50 times faster than COMSOL (579 s).

\begin{table}[h]
\renewcommand{\arraystretch}{1.3}
\caption{Parameters of the Propagation Modes for the Gradient Metasurface at Microwave Frequency}
\centering
\begin{tabular}{cccccc}
\hline
$\alpha$ &$\theta_{\alpha,k} (\textrm{deg})$&$\textbf{F}_{s}$&$|\tilde{\bar{G}}_{1,2}(\alpha,1)|^{2}$&$|\tilde{\bar{T}}_{1,N}(\alpha,1)|^{2}$\\
\hline
1 & 0.0000024  & -4.435190j &0.079939&0.000013\\

2 & 0.0000012  & 0.000000    &0.000000&0.248226\\

3 & 45.099347  & 0.000000    &0.000000&0.000000\\

4 & 45.099347  & 0.000000   &0.000000&0.000000\\

5 & 45.099834  & 0.000000   &0.297203&0.000000\\

6 & 45.099834  & 0.000000    &0.160337&0.000000\\

\hline
\end{tabular}
\label{example23}
\end{table}

\subsection{Extreme Ultraviolet (EUV) Lithography}

In microelectronics manufacturing, one critical technology is optical lithography used for circuit pattern production and reproduction. The research and development of lithography plays a leading role in the updating of each generation of integrated circuit technology. Here we consider an extreme ultraviolet (EUV) lithography model as a large-scale example \cite{Niu2017}, with large scatterers embedded in a stratified structure (including a 80-layer reflector in this case) acting as a light reflection mirror.

\begin{figure}[!t]
  \centering
  {
   \includegraphics[width=1.0\columnwidth,draft=false]{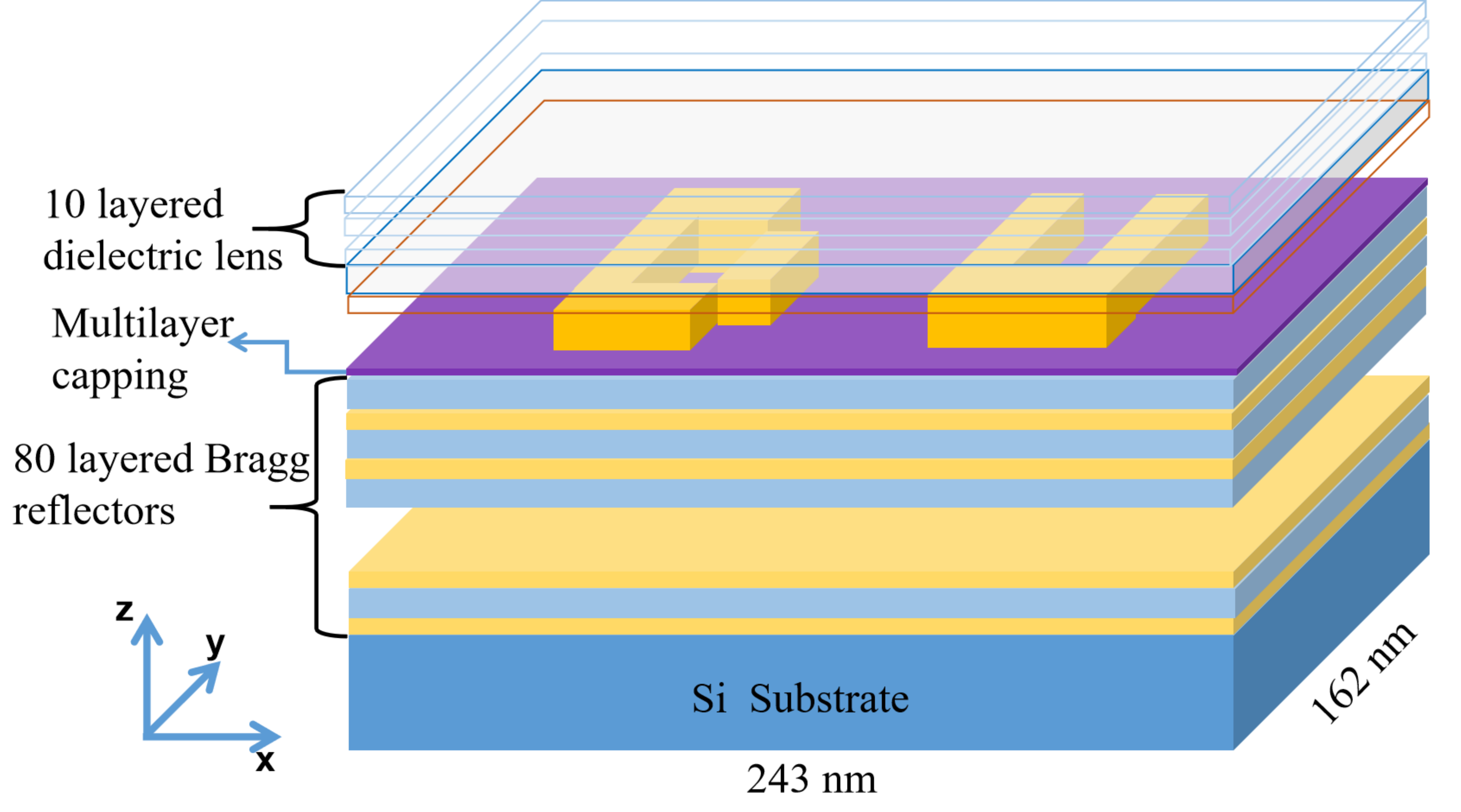}
   }
 \caption{Schematic drawing for the EUV lithography.}
\label{sketch4}
\end{figure}

As shown in Fig. \ref{sketch4}, we can see that including the first semi-infinite air layer and the last semi-infinite silicon substrate, the EUV lithography model has 96 layers; the dimensions of each principal cell are $243~\textrm{nm}\times162~\textrm{nm}$. The total thickness of this structure is 354.5 nm (excluding the first and last layers). When the working wavelength is $\lambda_{0}=13.5~\textrm{nm}$, if using the conventional numerical method (e.g., FEM, FDTD), a minimum computation domain with the size $18\lambda_{0}\times12\lambda_{0}\times27\lambda_{0}$ is to be solved. Therefore, the numerical simulation of the EUV lithography model is an electrically large problem. In addition, because the EUV lithography model has 96 films, it is difficult to build and partition this structure optimally in COMSOL. In order to compare with the SNMM method, we instead use finite element method (FEM) to solve the PBC waveguide problem (\ref{eq:6}) to obtain the eigenmodes for the NMM method, which is called the FNMM method.

From the top going downward, the parameters of the materials and geometries are listed as follows (also in \cite{Niu2017}):
\begin{itemize}
  \item  Air layer (infinitely thick)
  \item  Dielectric lens (5 bilayers: 3 nm-thick upper layer, $\epsilon_{r}=0.998$; 2 nm-thick lower layer, $\epsilon_{r}=0.854$ )
  \item Lithography filling background (thickness: 10 nm, $\epsilon_{r}=0.958$)
  \item Absorber capping (thickness: 12 nm, $\epsilon_{r}=0.7497-0.0296j$)
  \item  Lithography pattern (thickness: 27 nm, component width: 20 nm, $\epsilon_{r}=0.856-0.0807j$)
  \item Multilayer capping  (thickness: 2.5 nm, $\epsilon_{r}=0.75-0.0296j$)
  \item  Bragg reflector (40 bilayers: Si layer thickness, 4.17 nm with $\epsilon_{r}=0.998-0.000363j$; Mo layer thickness, 2.78 nm with $\epsilon_{r}=0.854-0.0119j$)
  \item Substrate (silicon, infinitely thick)
\end{itemize}
 In this example, the incident angle of the plane wave is $6^{0}$ and polarized with $\textbf{E}^{\textrm{inc}}\parallel \hat{x}$, i.e., $(\theta_{k}, \phi_{k}, \phi_{e})=(\pi/30,0,0)$. The 3rd order SNMM method with 320 eigenmodes and the 2nd order FNMM method with 320 eigenmodes are employed to this model, respectively. The DoFs are 1386 and 1240, respectively.

\begin{figure}[!t]
  \centering
  {
   \includegraphics[width=1.0\columnwidth,draft=false]{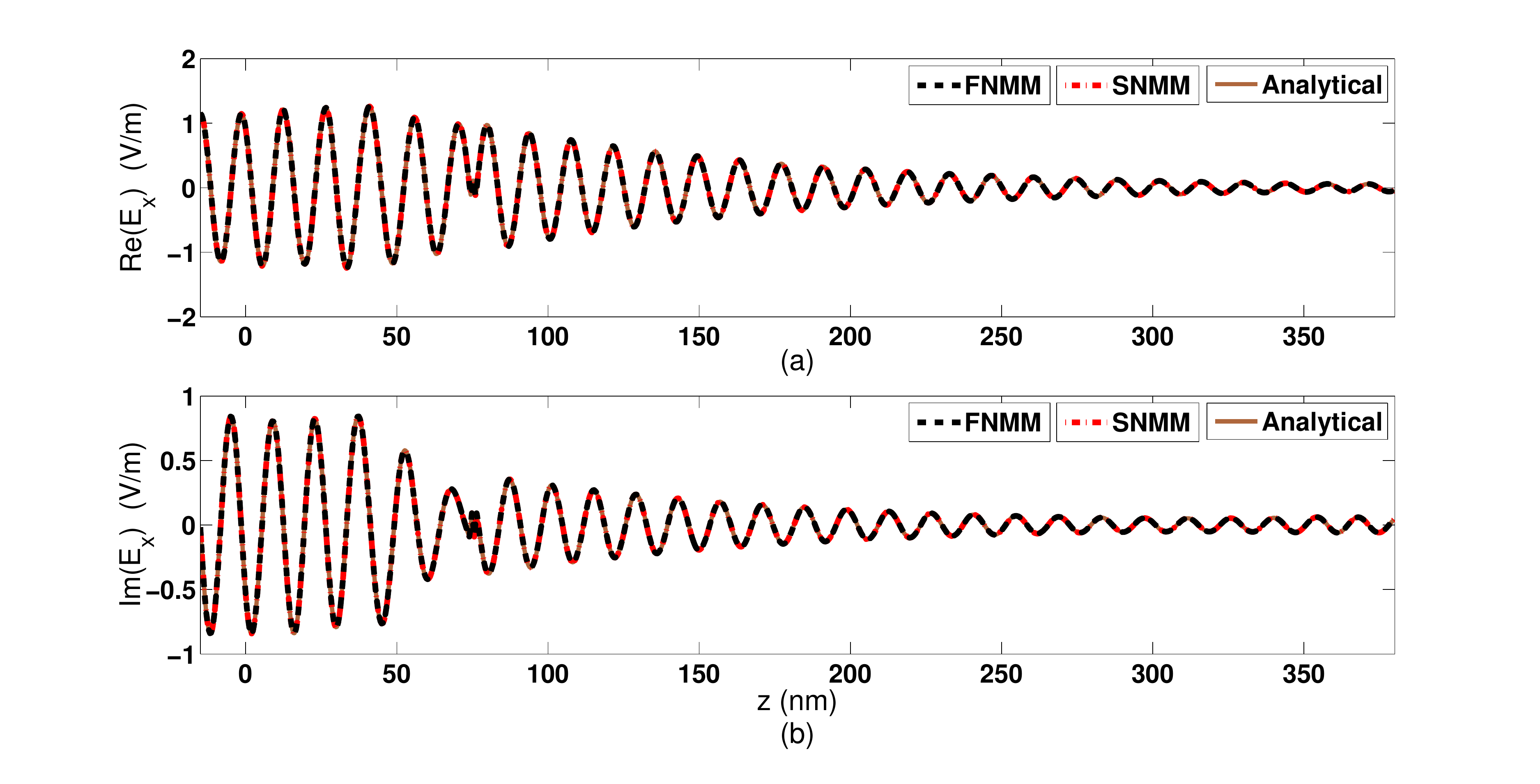}
   }
 \caption{Electric field component $E_{x}$ along the $z$-axis at $(x,y)=(0,0)$ for the lithography model with each layer filled with the homogeneous medium. (a) Real part of $E_{x}$. (b) Imaginary part of $E_{x}$.}
\label{tu31}
\end{figure}

 In order to verify the correctness and effectiveness of our methods, we first assume that the pattern layer is completely filled by the same material as the lithography pattern ($\epsilon_{r}=0.856-0.0807j$) (i.e., such that each layer is filled with the homogeneous isotropic medium), so that there is an analytical solution for this stratified medium. Fig. \ref{tu31} shows that the component $E_{x}$ obtained by the FNMM, the SNMM and the analytical solver can be well agreed. The relative error of the transverse electric field obtained by the FNMM and the analytical solver is $R\textbf{E}_{fa}=3.91\times 10^{-13}$; for the SNMM and the analytical solver, it follows that
 $R\textbf{E}_{sa}=3.74\times 10^{-13}$. The CPU time of the FNMM and the SNMM are 268.6 s and 371.4 s, respectively. In short, our two methods are correct and efficient for the stratified structures filled with the homogeneous isotropic medium.

\begin{figure}[!t]
  \centering
  {
   \includegraphics[width=1.0\columnwidth,draft=false]{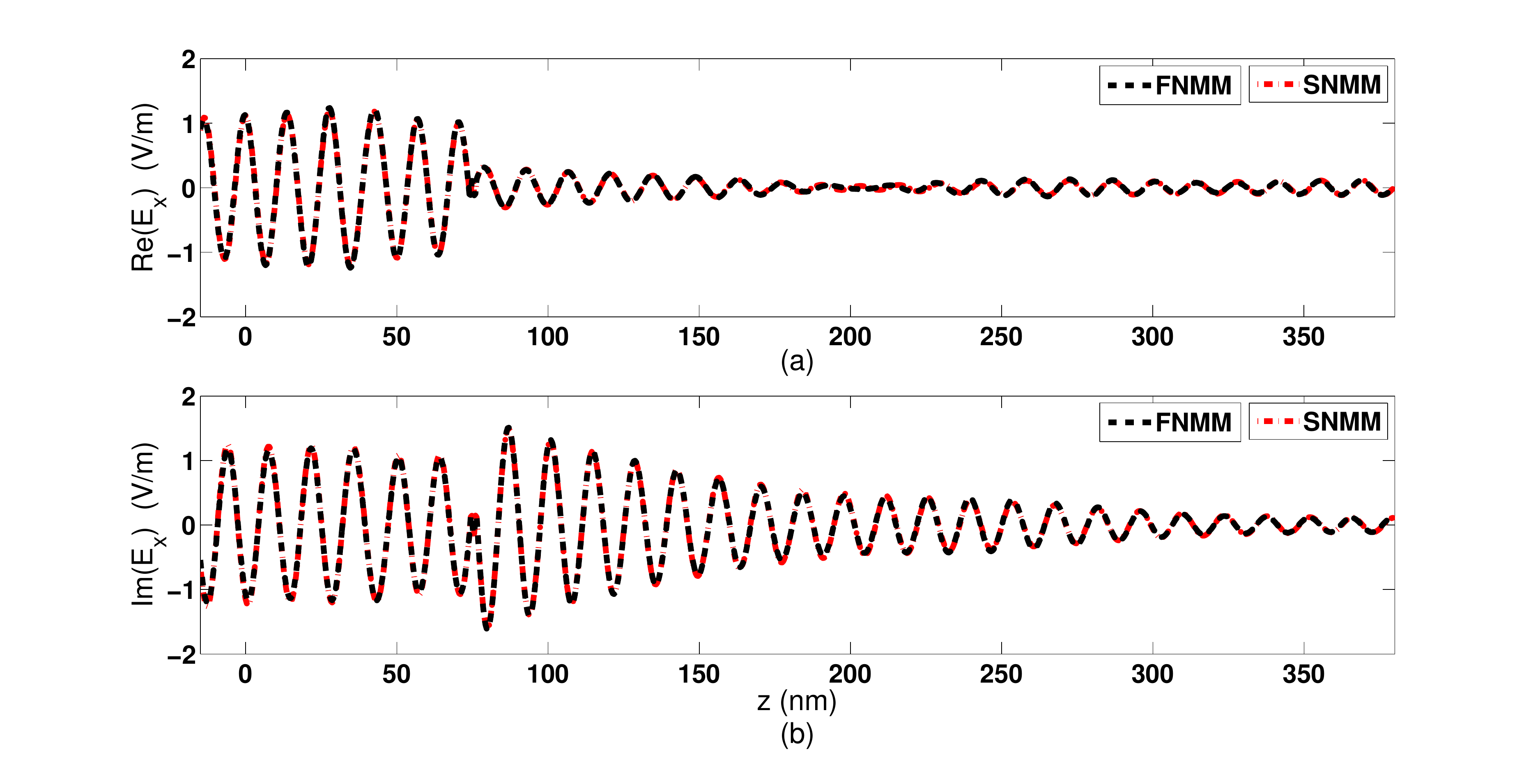}
   }
 \caption{Electric field component $E_{x}$ along the $z$-axis at $(x,y)=(0,0)$ for the lithography model with patterns $DU$. (a) Real part of $E_{x}$. (b) Imaginary part of $E_{x}$.}
\label{tu32}
\end{figure}

\begin{figure}[!t]
  \centering
  {
   \includegraphics[width=0.8\columnwidth,draft=false]{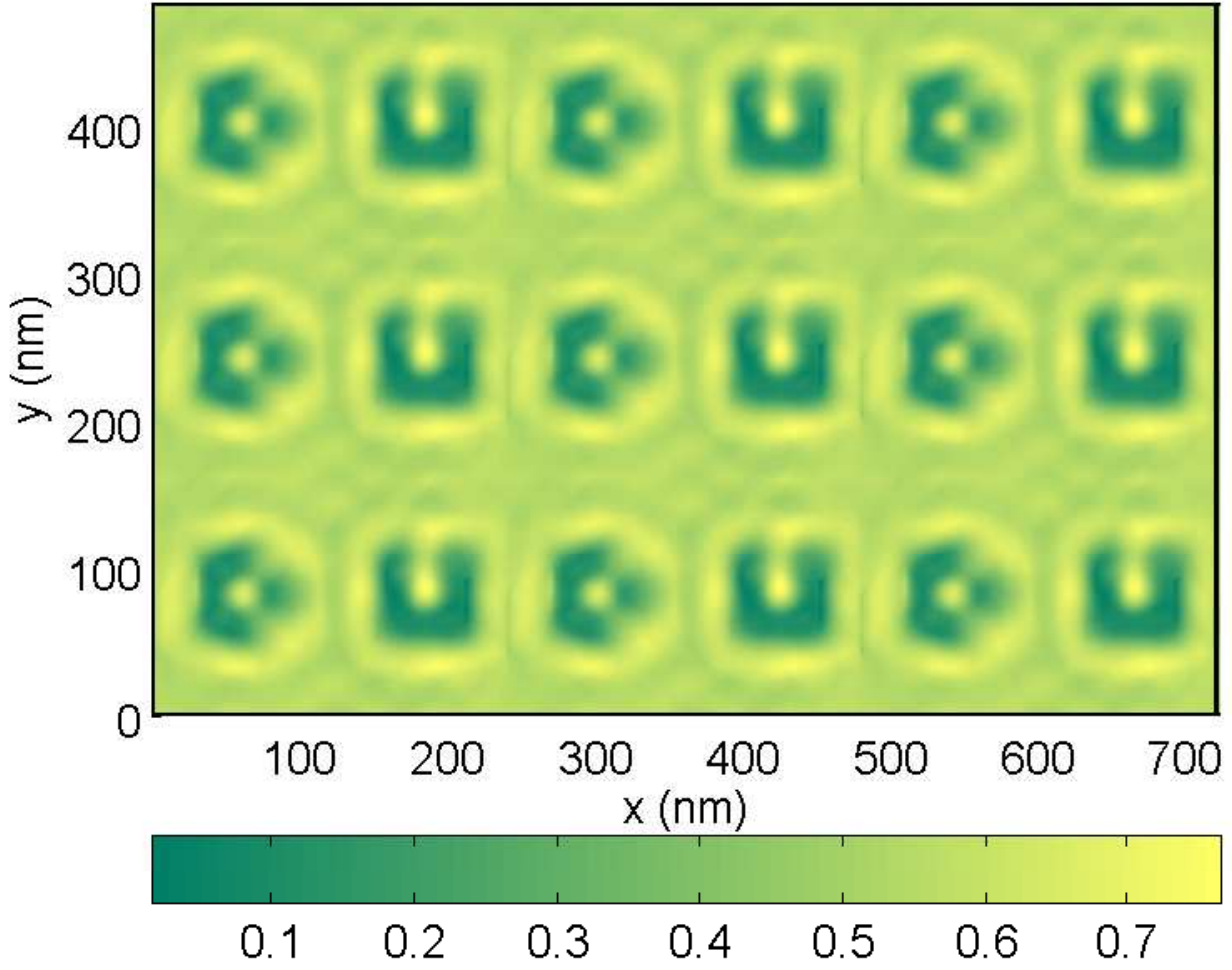}
   }
 \caption{Field distribution of the reflection field $\textbf{E}_{t}^{\textrm{ref}}$ (V/m) obtained by the SNMM method on the plane $z=12$ nm above the EUV lithography pattern.}
\label{tu33}
\end{figure}

\begin{figure}[!t]
  \centering
  {
   \includegraphics[width=0.8\columnwidth,draft=false]{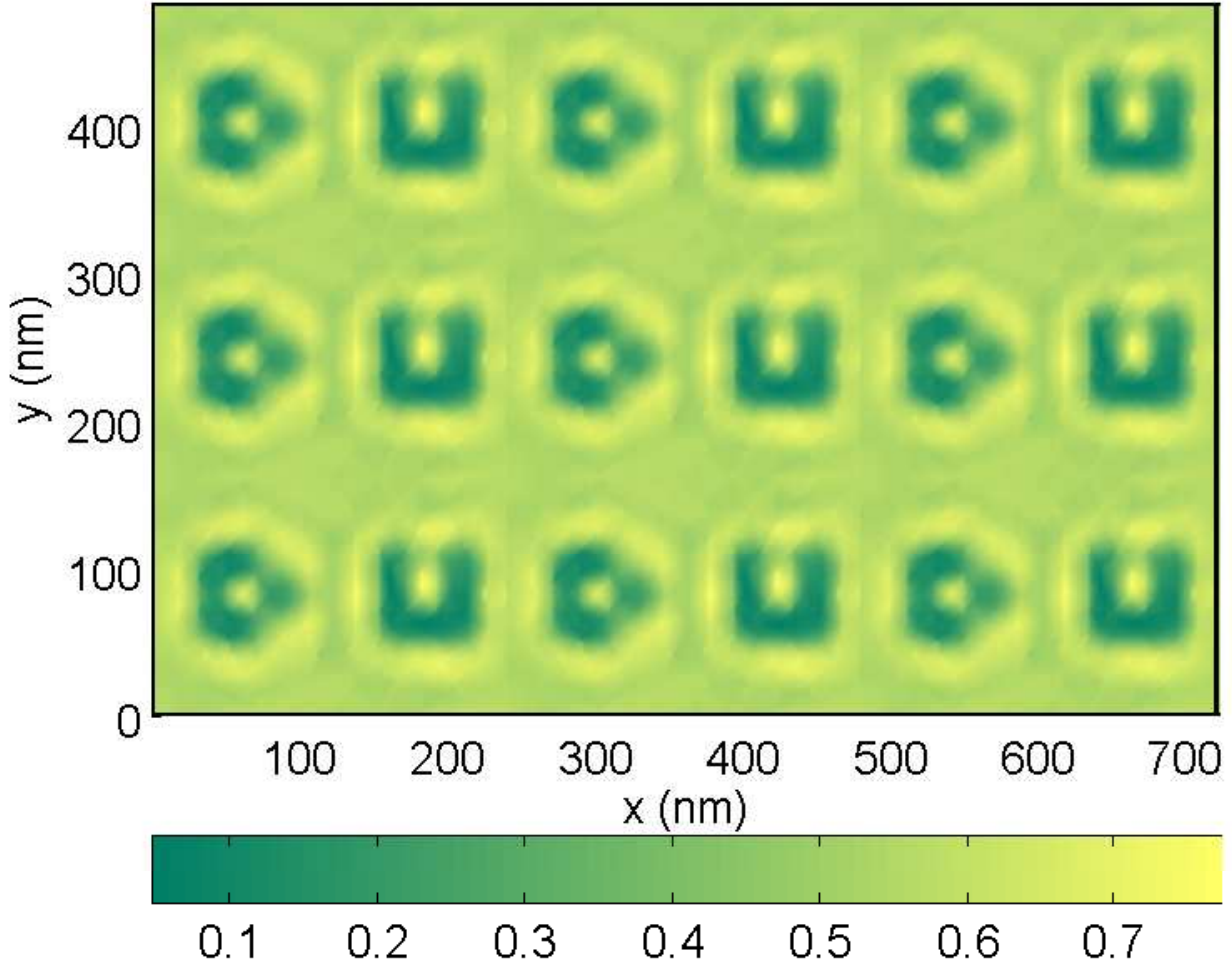}
   }
 \caption{Field distribution of the reflection field $\textbf{E}_{t}^{\textrm{ref}}$ (V/m) obtained by the FNMM method on the plane $z=12$ nm above the EUV lithography pattern.}
\label{tu34}
\end{figure}

 Finally, we turn to the EUV lithography model with the patterns $DU$, where only the pattern layer is filled the inhomogeneous media (air and $\epsilon_{r}=0.856-0.0807j$). Fig. \ref{tu32} shows that the component $E_{x}$ obtained by the FNMM and the SNMM  are well matched. The relative error of the transverse electric field obtained by the FNMM and the SNMM is $R\textbf{E}_{fs}=4.99\%$. The transverse reflection field $\textbf{E}_{t}^{\textrm{ref}}$ obtained by the SNMM method and the FNMM method are shown in Fig. \ref{tu33} and Fig. \ref{tu34}, respectively. Compared with the results of \cite{Niu2017}, our results are different owing to the width of the component of the lithography pattern (here they are set 20 nm) and the distance (here it is 62.25 nm) between the two patterns $D$ and $U$ may be different. The CPU time of the FNMM method and the SNMM method are 273.3 s and 374.6 s, respectively, which are less than that (15 min) in \cite{Niu2017}. In addition, our computation uses 3.78 GB memory, much less than 20 GB (65790 DOFs) in \cite{Niu2017}. As a result, the FNMM method and the SNMM method are highly accurate and efficient for EUV lithography simulations.

\subsection{A Simplified Lithography Model}
\begin{figure}[!t]
  \centering
  {
   \includegraphics[width=1.0\columnwidth,draft=false]{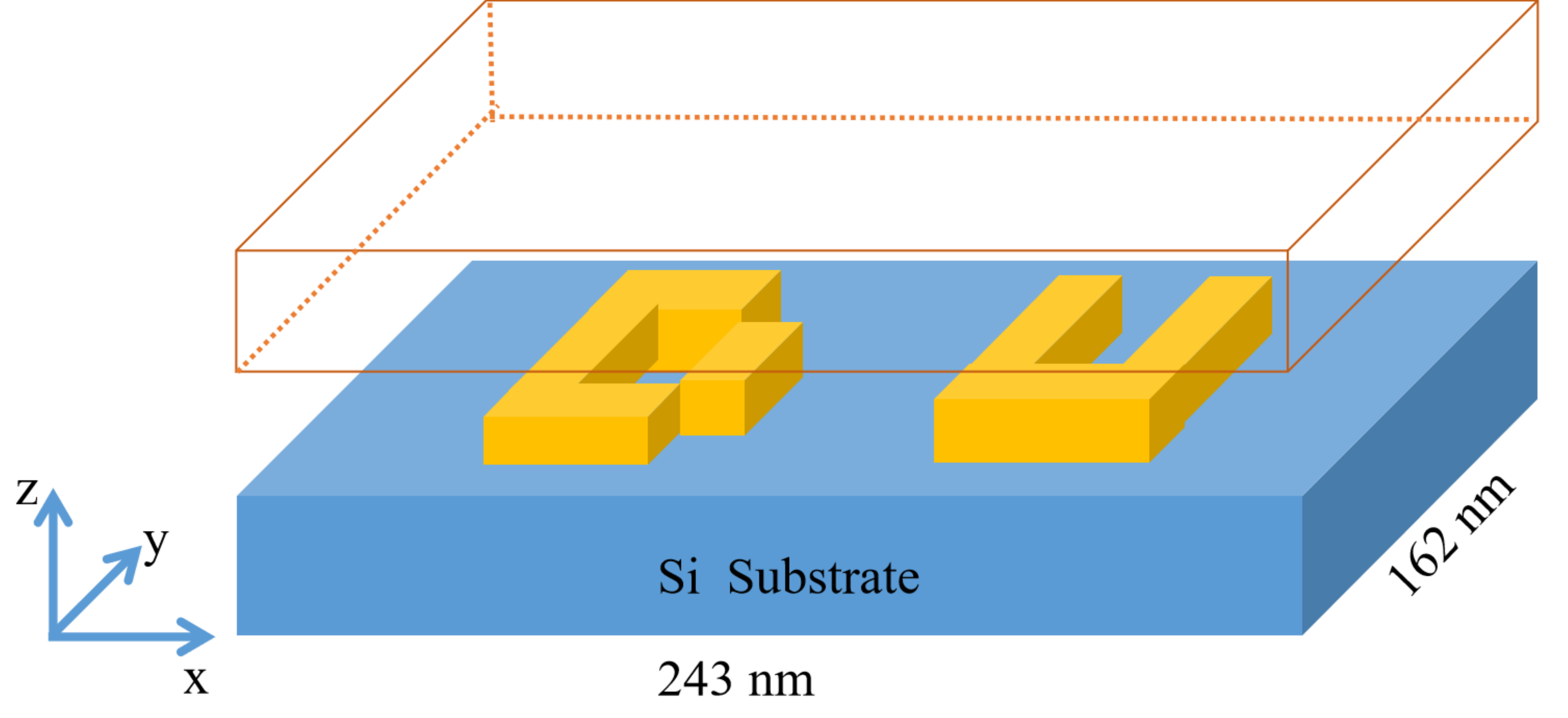}
   }
 \caption{Schematic drawing for the simplified lithography model.}
\label{sketch5}
\end{figure}

To verify that our SNMM method is more efficient than COMSOL, we discuss a simplified lithography model (see Fig. \ref{sketch5}) that is solvable by COMSOL. From the top downward, we assume that the first layer is the semi-infinite dielectric lens filled with $\epsilon_{r}=0.854$, the second layer is the 27 nm thick pattern ($\epsilon_{r}=0.856-0.0807j$) and the last layer is the semi-infinite silicon substrate ($\epsilon_{r}=0.998-0.00363j$). In order to verify that our methods are adapted to the plane waves with the different polarization and different incident angle, a plane wave with the operating wavelength $\lambda_{0}=13.5~\textrm{nm}$ is normally incident to this structure and the direction of polarization is along $\hat{y}$, i.e., $(\theta_{k}, \phi_{k}, \phi_{e})=(0,0,\pi/2)$. In COMSOL, the perfect matching layer (PML) absorbing boundary condition is used to truncate the first and the last layers, so that the thickness of the first and the last layers are 13.5 nm (a wavelength), and the PML is 7 nm thick at both ends of this structure. Therefore, COMSOL will compute a 3D structure with the size of about $18\lambda_{0}\times12\lambda_{0}\times5\lambda_{0}$.

\begin{figure}[!t]
  \centering
  {
   \includegraphics[width=1.0\columnwidth,draft=false]{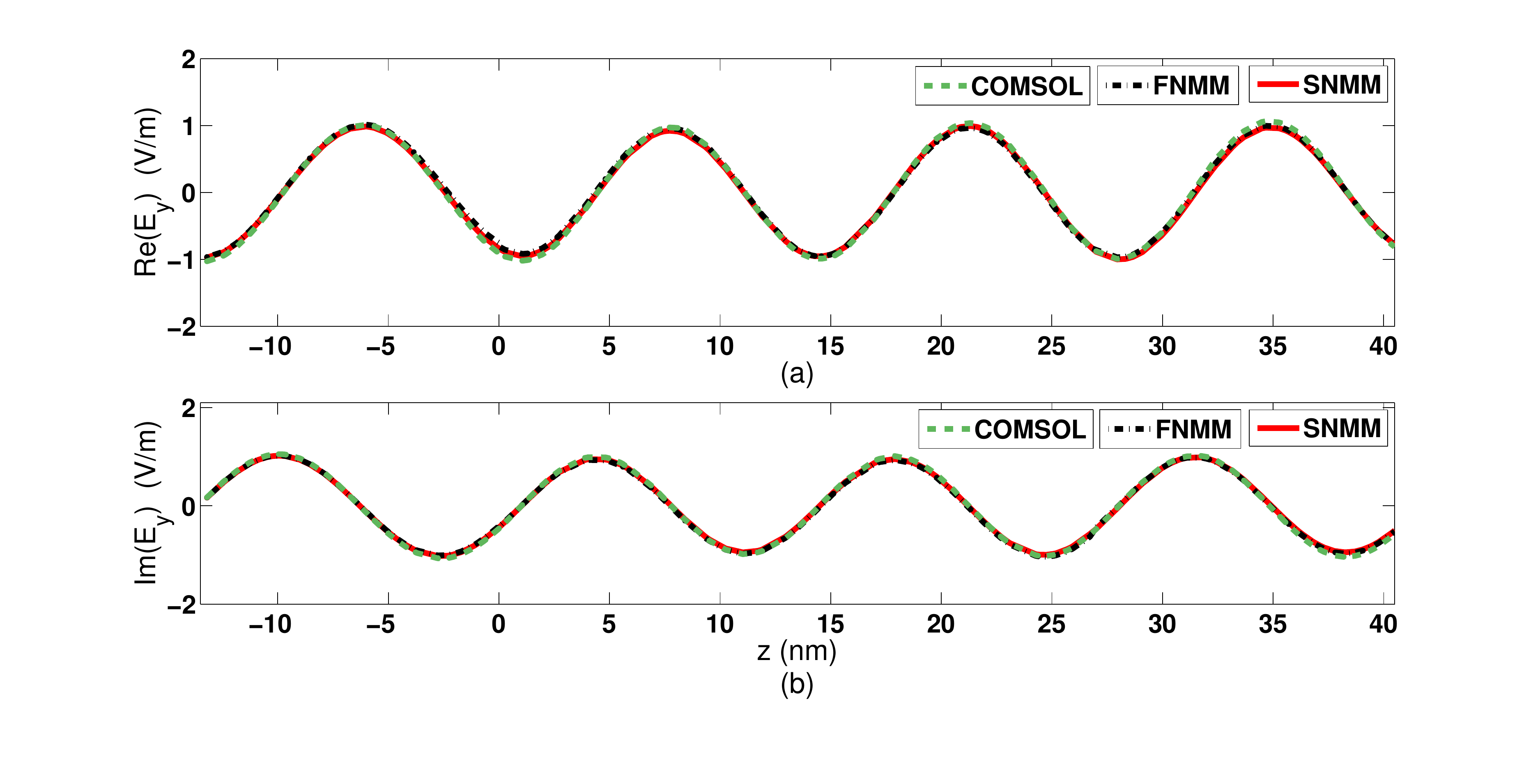}
   }
 \caption{Electric field component $E_{y}$ along the $z$-axis at $(x,y)=(0,0)$ for the simplified lithography model. (a) Real part of $E_{x}$. (b) Imaginary part of $E_{x}$.}
\label{tu41}
\end{figure}

The 4th order SNMM method with 1000 eigenmodes and the 2nd order FNMM method with 1000 eigenmodes are employed to solve this model, respectively. In COMSOL, the 3rd order FEM is employed to solve this 3D model. The computational costs of the three solvers are listed in Table \ref{example41}. We can see that COMSOL requires  87.94 times more memory than the SNMM method, and 36.13 times more than the FNMM method; the computational speeds of the SNMM method and the FNMM method are about 3.93 times faster than COMSOL; the DoF in COMSOL is about 1556.38 times larger than the SNMM method, and is 1775.42 times larger than the FNMM method. Fig. \ref{tu41} shows that the electric field component $E_{y}$ obtained by the SNMM method, the FNMM method and COMSOL are well matched. The relative errors of the transverse electric field obtained by the above three solvers are $R\textbf{E}_{fc}=6.14\%$, $R\textbf{E}_{sc}=5.89\%$ and
$R\textbf{E}_{fs}=3.98\%$, respectively. Therefore, our two methods are more accurate and efficient than COMSOL, and then they can be used to simulate EUV lithography model in the above example.
\begin{table}[!t]
\renewcommand{\arraystretch}{1.3}
\caption{Computational costs for the simplified lithography model}
\centering
\begin{tabular}{cccc}
\hline
Solver & Memory (GB)& CPU time (s) & DOF\\
\hline
COMSOL & 206.67 & 3239.00 & 3834909\\

SNMM   & 2.35 & 823.50 & 2464\\

FNMM & 5.72 &826.56 & 2160\\
\hline
\end{tabular}
\label{example41}
\end{table}

\section{Conclusion}

The spectral numerical mode-matching (SNMM) method is developed for the layered multi-region structure. The SNMM method is a semi-analytical solver. In the numerical part, the mixed spectral element method (MSEM) is used to solve the BPBC waveguide problem in the horizontal plane to obtain the Bloch eigenmodes. The highly accurate Bloch eigenmodes can be obtained because the MSEM is exponentially convergent and free of spurious modes. For the analytical part, along the vertical direction, the electric and magnetic fields are expressed as a summation of the Bloch eigenmodes, exponential factor consist of propagation constants, generalized reflection matrix, and excitation coefficients. These are obtained recursively with little computation cost.

The SNMM method is used to explore the characteristic parameters of the metasurface. The SNMM method is highly efficient to simulate metasurfaces, especially when some layers are thick compared to the wavelength. In addition, in order to verity that the SNMM method is also efficient for large-scale problems, it is also applied to simulate the extreme ultraviolet (EUV) lithography and a simplified lithography model. Numerical experiments indicate that the SNMM method is efficient and accurate for the metasurfaces and the lithography models.

%
%

\ifCLASSOPTIONcaptionsoff
  \newpage
\fi

\end{document}